%% file: main.tex
\documentclass[11pt,a4paper]{article}
\usepackage{jheppub}

\usepackage{tikz}
\usepackage[compat=1.1.0]{tikz-feynhand}
\usetikzlibrary{shapes.geometric}
\usetikzlibrary{positioning, intersections, calc, arrows.meta, math}
\usetikzlibrary{angles, quotes}
\usepackage{dynkin-diagrams}

\usepackage{amsthm}
\usepackage{bm}
\usepackage{bbm}
\usepackage{braket}
\usepackage{mathrsfs}
\usepackage{mathtools}
\usepackage{mleftright}

\usepackage{colortbl}
\usepackage{float}
\usepackage{array}
\usepackage{multirow}
\usepackage{adjustbox}

\numberwithin{equation}{section}
\numberwithin{figure}{section}
\numberwithin{table}{section}

\renewcommand{\Re}{\operatorname{Re}}

\preprint{TIT/HEP-710}

\title{The ODE/IM Correspondence between $C(2)^{(2)}$-type Linear Problems and 2d $\mathcal{N}=1$ SCFT}
\author{Naozumi Tanabe}

\affiliation{Department of Physics, Institute of Science Tokyo,\\
Tokyo, 152-8551, Japan}

\emailAdd{naotanabe.hep@gmail.com}

\abstract{
  We study the ODE/IM correspondence between the linear problem associated with the supersymmetric affine Toda field equation for the twisted affine Lie superalgebra $C(2)^{(2)} = \mathfrak{osp}(2|2)^{(2)}$ and two-dimensional $\mathcal{N}=1$ superconformal field theories (SCFTs). 
  On the ODE side, we introduce a boundary condition more suitable for the conformal limit and the subsequent WKB analysis and diagonalize the resulting Lax operator. 
  This leads to a WKB expansion from which we extract the WKB periods and semi-local conserved quantities up to tenth order. 
  On the IM side, we compute the eigenvalues of the local integrals of motion on the cylinder in both the Neveu--Schwarz and Ramond sectors of 2d $\mathcal{N}=1$ SCFTs. 
  We then compare the two sides and verify, up to sixth order, that the WKB periods coincide with the eigenvalues of the local integrals of motion for highest-weight states in the Neveu--Schwarz sector. 
}


\arxivnumber{2604.14899}

\begin{document}
\mleftright
\maketitle

\input{Introduction}
\input{super_Toda}
\input{C22_linear}
\input{SCFT}
\input{ODEIM}
\input{Conclusion}

\acknowledgments
The author is grateful to Prof.~Katsushi Ito for valuable discussions and helpful comments on the manuscript. 
The author also thanks Wataru Kono for useful advice on the calculations. 
The author used ChatGPT to assist with Mathematica coding and with improving the English expression of parts of the manuscript. 
All outputs were checked and revised by the author, who takes full responsibility for the content of this work. 

\appendix
\input{WKB_period}
\input{conf_map}

\bibliographystyle{JHEP}
\bibliography{C22}
\end{document}

%% file: Introduction.tex
\section{Introduction}
\label{sec:introduction}
Two-dimensional conformal field theories (CFTs) possess rich integrable structures, encoded in infinite families of mutually commuting local integrals of motion (IoMs). 
These integrable structures were first studied systematically by Bazhanov, Lukyanov, and Zamolodchikov (BLZ) \cite{bazhanov1996,bazhanov1997,bazhanov1999}. 
In their work, local and semi-local IoMs were constructed for Virasoro conformal field theories, or equivalently for the quantum KdV hierarchy. 
Since then, this framework has been extended to a variety of models, including theories with $\mathcal{W}$-algebra symmetry \cite{fioravanti1996,bazhanov2002,litvinov2021,ashok2024,ito2025a,kudrna2025,ide2026}
and the super-Virasoro case associated with the super-KdV hierarchy \cite{kulish2004,kulish2005}. 
\par
An important further development was the discovery that these integrable structures admit an alternative description in terms of ordinary differential equations. 
This led to the Ordinary Differential Equation/Integrable Model (ODE/IM) correspondence, which relates spectral problems of differential equations to integrable models in two dimensions \cite{dorey1999,dorey1999a,bazhanov1997a,bazhanov2003}\footnote{For reviews and introductory accounts of the ODE/IM correspondence, see \cite{dorey2007,ito2025}. }. 
In particular, the BLZ approach was enriched by the observation that the eigenvalues of local and semi-local IoMs can be encoded in the spectral data of suitable ODEs. 
This correspondence provides a powerful link between conformal field theory, quantum integrable models, and the spectral theory of differential operators. 
\par
It is therefore of considerable interest to study the relation between the IoMs in CFT and the WKB periods of the corresponding ODEs. 
For CFTs with $W$-currents, the ODE/IM correspondence for higher-order differential equations has been studied extensively, in particular from the viewpoint of integral equations \cite{dorey2000,suzuki1999,dorey2000a,dorey2007a}. 
In particular, $\mathcal{W} \, \widehat{\mathfrak{g}}$-CFTs are expected to correspond to linear problems obtained from the affine Toda field equations associated with $\widehat{\mathfrak{g}}^\vee \,$\footnote{$\; \widehat{\mathfrak{g}}^\vee$ denotes the Langlands dual of $\widehat{\mathfrak{g}}$. } after a suitable conformal modification \cite{bazhanov2002,ito2014}; 
see also \cite{locke2015}. 
The correspondence for simply-laced affine Lie algebras has been further investigated in \cite{bazhanov2002,ashok2024,ito2023a,ito2025a,kudrna2025,ide2026}; 
see also \cite{zhu2025a}. 
\par
A particularly relevant supersymmetric example arises in the ODE/IM correspondence for the $SU (2)$ coset models \cite{lukyanov2007,dorey2007a} 
\begin{equation}
  \frac{SU (2)_K \times SU (2)_L}{SU (2)_{K + L}} . 
\end{equation}
For $K = 2$, the chiral algebra is enhanced to the $\mathcal{N} = 1$ superconformal algebra. 
The corresponding ODE/IM relation was studied in \cite{babenko2017}; see also \cite{babenko2019a}, where the WKB periods were related to quantities obtained from the Suzuki equation \cite{suzuki1999a,dunning2003}. 
\par
On the other hand, a supersymmetric affine Toda type formulation of the ODE side was developed in \cite{ito2022a}, where $\mathcal{N} = 1$ supersymmetric affine Toda field equations associated with affine Lie superalgebras with purely odd simple roots were constructed together with the corresponding linear problems. 
In particular, the $C (2)^{(2)} = \mathfrak{osp} (2 | 2)^{(2)}$ case provides the supersymmetric sinh-Gordon equation and leads, after taking the conformal limit, to the ODE relevant to the $\mathcal{N} = 1$ superconformal theory. 
\par
On the SCFT side, the integrable structure of 2d $\mathcal{N} = 1$ superconformal field theory has been studied from the viewpoint of the super-KdV hierarchy and related quantum inverse scattering constructions \cite{kulish2004,kulish2005,mathieu1990,yamanaka1988,alfimov2015,chistyakova2022}. 
These works clarify the structure of local and semi-local IoMs in SCFT, and provide the natural IM counterpart of the supersymmetric ODE data. 
\par
Taken together, these developments strongly suggest a supersymmetric ODE/IM correspondence. 
However, a direct comparison based on the diagonalization of the full $C (2)^{(2)}$-type linear problem and the explicit NS-sector eigenvalues of local IoMs has remained incomplete. 
In the present paper, we address this problem by diagonalizing the $C (2)^{(2)}$-type linear problem without taking the bosonic limit, computing the resulting local WKB periods, and comparing them with the NS-sector eigenvalues of local IoMs in the $\mathcal{N} = 1$ SCFT up to sixth order. 
\par
This paper is organized as follows. 
In Section~\ref{sec:super-Toda}, we review the $\mathcal{N} = 1$ supersymmetric affine Toda field equations following \cite{ito2022a}, and formulate the corresponding linear problem without taking the bosonic limit. 
We then introduce the associated linear problem by taking the conformal and light-cone limits of the supersymmetric affine Toda equation. 
In Section~\ref{sec:linear_problem}, we specialize to the $C (2)^{(2)}$ case, explicitly diagonalize the linear problem, and compute the corresponding WKB periods. 
In Section~\ref{sec:SCFT}, we review the $\mathcal{N} = 1$ superconformal field theory. 
We then present the free-field realization of the super-Virasoro algebra and compute the eigenvalues of the IoMs on the cylinder. 
Finally, in Section~\ref{sec:ODE/IM}, we compare the WKB periods with the eigenvalues of the IoMs on the cylinder and verify the ODE/IM correspondence. 
In Appendix~\ref{sec:higher_WKB}, we list higher-order WKB periods. 
In Appendix~\ref{sec:conf_map}, we derive normal ordering formulae on the cylinder, in particular for anti-periodic operators, and present several illustrative examples of zero mode computations. 

%% file: super_Toda.tex
\section{Supersymmetric affine Toda field equations}
\label{sec:super-Toda}
In this section, we review the $\mathcal{N} = 1$ supersymmetric affine Toda field equations associated with affine Lie superalgebras admitting a purely odd simple root system, and derive the corresponding linear problems. 
Our presentation follows the framework of \cite{ito2022a}, but we formulate the linear problem without taking the bosonic limit, in order to prepare for the subsequent WKB analysis of the full $C (2)^{(2)} = \mathfrak{osp} (2 | 2)^{(2)}$ system. 
The specialization to the $C (2)^{(2)}$ case will be carried out in the following sections. 
\par
Let $\widehat{\mathfrak{g}}$ be an affine Lie superalgebra with simple roots $\{ \alpha_0 , \alpha_1 , \dots , \alpha_r \}$, all of which are odd. 
In this paper, we restrict ourselves to the case of twisted affine Lie superalgebras with $m = 2$. 
\par
Now we introduce the $\mathcal{N} = 1$ supersymmetric affine Toda field equation associated with an affine Lie superalgebra $\widehat{\mathfrak{g}}$ \cite{olshanetsky1983}. 
To realize $\mathcal{N} = 1$ supersymmetry, we introduce the $\mathcal{N} = 1$ superspace with complex coordinates $\left( z , \theta , \bar{z} , \bar{\theta} \, \right)$, where $\theta , \bar{\theta}$ are Grassmann variables. 
The supercovariant derivatives are defined by 
\begin{align}
  D &\coloneqq \frac{\partial}{\partial \theta} + \theta \, \partial_z , \quad 
  \bar{D} \coloneqq \frac{\partial}{\partial \bar{\theta}} + \bar{\theta} \, \partial_{\bar{z}} , 
  \label{eq:covariant_superderivative} \\
  D^2 &= \partial_z , \quad \bar{D}^2 = \partial_{\bar{z}} , \quad 
  \left\{ D , \bar{D} \right\} = 0 . 
  \label{eq:covariant_superderivative2}
\end{align}
We also introduce a $r$-component scalar superfield by 
\begin{equation}
  \varPhi \left( z , \bar{z} , \theta , \bar{\theta} \, \right) 
  = \phi \left( z , \bar{z} \right) + i \theta \, \eta \left( z , \bar{z} \right) 
  + i \bar{\theta} \, \bar{\eta} \left( z , \bar{z} \right) + \theta \bar{\theta} \, F \left( z , \bar{z} \right) , 
  \label{eq:sfld}
\end{equation}
where $\phi \left( z , \bar{z} \right)$ and $F \left( z , \bar{z} \right)$ are bosonic fields, while $\eta \left( z , \bar{z} \right)$ and $\bar{\eta} \left( z , \bar{z} \right)$ are fermionic fields. 
\par
The action of the $\mathcal{N} = 1$ supersymmetric Toda field theory for a Lie superalgebra $\mathfrak{g}$ with purely odd simple roots is defined by 
\begin{equation}
  S [\varPhi] = \int d^2 z \, d^2 \theta \left[
    \frac{1}{2} D \varPhi \cdot \bar{D} \varPhi 
    - \frac{m^2}{\beta^2} \, \sum_{i = 1}^{r} e^{\beta \, \alpha_i \cdot \varPhi} 
  \right] , 
  \label{eq:action_Toda}
\end{equation}
where $\beta$ is a coupling constant, $m$ is a mass parameter, and $\alpha_i$ are the simple roots of $\mathfrak{g}$ \cite{evans1991,evans1992,komata1991,delduc1992}. 
The symbol ``$\; \cdot \;$'' denotes the inner product in the root space. 
The action of the affine Toda field theory for an affine Lie superalgebra $\widehat{\mathfrak{g}}$ is obtained by adding the affine root contribution $- \frac{m^2}{\beta^2} \, e^{\beta \alpha_0 \cdot \varPhi}$. 
Accordingly, the affine Toda field equation takes the form 
\begin{equation}
  D \bar{D} \varPhi + \frac{m^2}{\beta} \sum_{i = 1}^{r} \alpha_i \, e^{\beta \, \alpha_i \cdot \varPhi} 
  + \frac{m^2}{\beta} \alpha_0 \, e^{\beta \, \alpha_0 \cdot \varPhi} = 0 . 
  \label{eq:Toda_eq}
\end{equation}

\subsection{Modified supersymmetric affine Toda field equations}
\label{subsec:super-Toda_eq}
The Toda field equation \eqref{eq:Toda_eq} without the affine root term is invariant under the superconformal transformation 
\begin{equation}
  \begin{aligned}
  z' (z , \theta) &= f (z) + \theta \, \chi (z) \, g (z) , \quad 
  \theta' = \chi (z) + \theta \, g (z) , \\
  \bar{z}' \left( \bar{z} , \bar{\theta} \, \right) 
  &= \bar{f} \left( \bar{z} \right) 
  + \bar{\theta} \, \bar{\chi} \left( \bar{z} \right) \, \bar{g} \left( \bar{z} \right) , \quad 
  \bar{\theta}' = \bar{\chi} \left( \bar{z} \right) 
  + \bar{\theta} \, \bar{g} \left( \bar{z} \right) , \\
  \varPhi' \left( z , \theta , \bar{z} , \bar{\theta} \, \right) 
  &= \varPhi \left( z' , \theta' , \bar{z}' , \bar{\theta}' \right) 
  + \frac{\mu}{\beta} \log \left( D \theta' \, \bar{D} \bar{\theta}' \right) ,
  \end{aligned}
  \label{eq:superconf_transf}
\end{equation}
where $f (z) ,\, g (z) ,\, \bar{f} \left( \bar{z} \right)$ and $\bar{g} \left( \bar{z} \right)$ are bosonic functions, while $\chi (z)$ and $\bar{\chi} \left( \bar{z} \right)$ are fermionic functions. 
They satisfy 
\begin{equation}
  g (z)^2 = \partial_z f (z) + \chi (z) \, \partial_z \chi (z) , \quad
  \bar{g} \left( \bar{z} \right)^2 = \partial_{\bar{z}} \bar{f} \left( \bar{z} \right) 
  + \bar{\chi} \left( \bar{z} \right) \, \partial_{\bar{z}} \bar{\chi} \left( \bar{z} \right) . 
  \label{eq:superconf_cond}
\end{equation}
Here $\mu = \frac{1}{2} \sum \limits_{i = 1}^{r} \mu_i$, where $\mu_i$ are the fundamental weights satisfying $\mu_i \cdot \alpha_j = \delta_{i j}$. 
Under this transformation, the affine Toda field equation \eqref{eq:Toda_eq} is mapped to 
\begin{equation}
  D \bar{D} \varPhi 
  + \frac{m^2}{\beta} \sum_{i = 1}^{r} \alpha_i \exp (\beta \alpha_i \cdot \varPhi) 
  + \frac{m^2}{\beta} \, P (z , \theta) \, \bar{P} \left( \bar{z} , \bar{\theta} \, \right) \, 
  \alpha_0 \exp (\beta \alpha_0 \cdot \varPhi) = 0 , 
  \label{eq:mod_affine_Toda_eq}
\end{equation}
where 
\begin{equation}
  \begin{aligned}
    P (z , \theta) &\coloneqq \left( D \theta' \right)^{\! h} 
    = g (z)^h + \theta \, h \, g (z)^{h - 1} \, \partial_z \chi (z) , \\
    \bar{P} \left( \bar{z} , \bar{\theta} \, \right) 
    &\coloneqq \left( \bar{D} \bar{\theta}' \right)^{\! h} 
    = \bar{g} \left( \bar{z} \right)^h + \bar{\theta} \, h \, \bar{g} \left( \bar{z} \right)^{h - 1} \, 
    \partial_{\bar{z}} \bar{\chi} \left( \bar{z} \right) . 
  \end{aligned}
  \label{eq:superspace_potential}
\end{equation}
The exponent $h$ is determined by $\mu \cdot \alpha_0 = - \frac{h - 1}{2}$. 
If the extended root is expanded as $\alpha_0 = - \sum \limits_{i = 1}^{r} n_i \alpha_i$, then 
\begin{equation}
  h = \sum_{i = 1}^{r} n_i + 1 , 
\end{equation}
namely the Coxeter number of $\mathfrak{g}$. 
The equation \eqref{eq:mod_affine_Toda_eq} is no longer manifestly supersymmetric in the coordinates $(z,\bar{z})$, because the affine root term now carries the explicit factors $P (z , \theta)$ and $\bar{P} \left( \bar{z} , \bar{\theta} \, \right)$. 
Nevertheless, it is related to the supersymmetric equation \eqref{eq:Toda_eq} through the superconformal transformation \eqref{eq:superconf_transf}. 

\subsection{Super-Lax operator}
\label{subsec:super-Lax}
We now rewrite the modified affine Toda field equation \eqref{eq:mod_affine_Toda_eq} as a flatness condition of a super-Lax pair. 
Let $E_{\pm \alpha_i} \; (i = 0 , 1 , \dots , r)$ denote the root generators of $\widehat{\mathfrak{g}}$, and let $H$ denote the Cartan generators. 
They satisfy the following defining (anti-)commutation relations: 
\begin{equation}
  \{ E_{\alpha_i} , E_{- \alpha_i} \} = \varepsilon_i \, \alpha_i \cdot H , \quad 
  [\alpha_i \cdot H , E_{\pm \alpha_j}] 
  = \pm (\alpha_i \cdot \alpha_j) E_{\pm \alpha_j} , 
  \label{eq:superalgebra}
\end{equation}
where $\varepsilon_i$ is a normalization constant chosen as $\pm 1$. 
\par
We introduce the super-Lax operators 
\begin{equation}
  \begin{aligned}
  \mathcal{L}_{\mathrm{F}} &\coloneqq D - A_\theta = D - \beta \, D \varPhi \cdot H 
  - m e^\lambda \left[ \sum_{i = 1}^{r} E_{\alpha_i} + P (z , \theta) \, E_{\alpha_0} \! \right] , \\
  \bar{\mathcal{L}}_{\mathrm{F}} &\coloneqq \bar{D} - A_{\bar{\theta}} = \bar{D} 
  + m e^{- \lambda} \left[ \sum_{i = 1}^{r} e^{\beta \, \alpha_i \cdot \varPhi} \varepsilon_i E_{- \alpha_i} 
  + \bar{P} \left( \bar{z} , \bar{\theta} \, \right) \, 
  e^{\beta \alpha_0 \cdot \varPhi} \varepsilon_0 E_{- \alpha_0} \! \right] , 
  \end{aligned}
  \label{eq:super-Lax_F}
\end{equation}
where $\lambda$ is the spectral parameter. 
Then the modified affine Toda field equation \eqref{eq:mod_affine_Toda_eq} is equivalent to the flatness condition 
\begin{equation}
  F_{\theta \bar{\theta}} = D A_{\bar{\theta}} + \bar{D} A_\theta - \{ A_\theta , A_{\bar{\theta}} \} = 0 . 
  \label{eq:flatness_cond}
\end{equation}
Equivalently, it appears as the compatibility condition for the supersymmetric linear problems 
\begin{equation}
  \mathcal{L}_{\mathrm{F}} \varPsi \left( z , \bar{z} , \theta , \bar{\theta} \, \right) = 0 , \quad 
  \bar{\mathcal{L}}_{\mathrm{F}} \varPsi \left( z , \bar{z} , \theta , \bar{\theta} \, \right) = 0 . 
  \label{eq:SUSY_linear_problem}
\end{equation}
\par
Following the approach of Lukyanov and Zamolodchikov \cite{lukyanov2010}, we now consider the holomorphic part of the supersymmetric linear problem \eqref{eq:SUSY_linear_problem} in components. 
Writing $\varPsi (z , \theta) = \varPsi_0 (z) + i \theta \, \varPsi_1 (z) ,\; 
D \varPhi (z , \theta) = i \eta (z) + \theta \, \partial_z \phi (z) ,\; 
P (z , \theta) = p_0 (z) + \theta \, p_1 (z)$ and suppressing the anti-holomorphic part, the first equation in \eqref{eq:SUSY_linear_problem} becomes 
\begin{equation}
  \begin{aligned}
    \partial_z \varPsi_0 
    - \beta \, \partial_z \phi \cdot H \, \varPsi_0 
    - m e^\lambda \, p_1 (z) \, E_{\alpha_0} \varPsi_0 
    - i \left\{ \! m e^\lambda \left[ \sum_{i = 1}^{r} E_{\alpha_i} + p_0 (z) \, E_{\alpha_0} \! \right] 
    - i \beta \, \eta \cdot H \! \right\} \varPsi_1 &= 0 , \\
    i \varPsi_1 - i \beta \, \eta \cdot H \, \varPsi_0 
    - m e^\lambda \left[ \sum_{i = 1}^{r} E_{\alpha_i} + p_0 (z) \, E_{\alpha_0} \! \right] \varPsi_0 &= 0 . 
  \end{aligned}
  \label{eq:holomorphic_linear_problem}
\end{equation}
Eliminating the fermionic component $\varPsi_1$, we arrive at the bosonic linear problem $\mathcal{L}_{\mathrm{B}} \varPsi_0 = 0$ with 
\begin{align}
  \mathcal{L}_{\mathrm{B}} = \partial_z 
  &- \beta \, \partial_z \phi \cdot H 
  - m e^\lambda \, p_1 (z) \, E_{\alpha_0} \notag \\
  &- \left\{ \! m e^\lambda \left[ \sum_{i = 1}^{r} E_{\alpha_i} + p_0 (z) \, E_{\alpha_0} \! \right] 
  - i \beta \, \eta \cdot H \! \right\} \! \left\{ \! m e^\lambda \left[ \sum_{i = 1}^{r} E_{\alpha_i} 
  + p_0 (z) \, E_{\alpha_0} \! \right] + i \beta \, \eta \cdot H \! \right\} . 
  \label{eq:bosonic-Lax}
\end{align}
After taking the conformal limit, this becomes a linear problem in a single holomorphic variable, to which the ODE/IM correspondence will be applied. 

\subsection{Conformal limit of the linear problem}
\label{subsec:conf_lim}
We now consider the conformal and light-cone limit of the supersymmetric linear problem \eqref{eq:SUSY_linear_problem}, in which the anti-holomorphic dependence becomes negligible. 
For this purpose, we choose the bosonic part of the potential to be of monomial type,
\begin{equation}
  p_0 (z) = z^{\frac{h M}{2}} - s^{\frac{h M}{2}} , \quad 
  \bar{p}_0 \left( \bar{z} \right) 
  = \bar{z}^{\frac{h M}{2}} - \bar{s}^{\frac{h M}{2}} . 
  \label{eq:potential}
\end{equation}
Here $M$ is a positive real parameter satisfying $M > \frac{1}{h - 1}$, while $s$ is an arbitrary parameter. 
\par
To specify the conformal limit, we fix the asymptotic behavior of the superfield near infinity and near the origin. 
As $|z| \to \infty$, the potential term dominates, and the leading behavior of $\varPhi \left( z , \bar{z} , \theta , \bar{\theta} \, \right)$ is 
\begin{equation}
  \varPhi \left( z , \bar{z} , \theta , \bar{\theta} \, \right) 
  = \frac{M \mu}{\beta} \log \left( z \bar{z} \right) + \cdots . 
  \label{eq:sfld_expansion_inf}
\end{equation}
Near the origin, we impose the boundary condition 
\begin{equation}
  \varPhi \left( z , \bar{z} , \theta , \bar{\theta} \, \right) 
  = \frac{l}{\beta} \log \left( z \bar{z} \right) 
  + \theta \, \frac{\xi}{\beta z} + \bar{\theta} \, \frac{\bar{\xi}}{\beta \bar{z}} + \cdots . 
  \label{eq:sfld_expansion_0}
\end{equation}
Here the coefficient $l$ characterizes the leading bosonic logarithmic behavior near the origin and plays a role analogous to an angular-momentum parameter in the associated ODE. 
Similarly, the Grassmann-odd parameters $\xi$ and $\bar{\xi}$ arise from the leading singular terms of the fermionic components $\eta$ and $\bar{\eta}$. 
\par
We now pass to the conformal and light-cone limit of the linear problem. 
For this purpose, we first perform the gauge transformation
\begin{equation}
  A_\theta \to A^U_\theta = U A_\theta U^{-1} - U \, D U^{-1} , \quad 
  \varPsi \to U \varPsi , 
  \label{eq:gauge_transf}
\end{equation}
with $U = \exp \left[ \frac{2}{h} \log P (z , \theta) \, \mu \cdot H \right]$. 
This transforms the $P (z , \theta)$-dependence of the super-Lax operator into a form better suited to the scaling limit. 
The resulting connection is
\begin{equation}
  A^U_\theta = \beta \, D \varPhi \cdot H - \frac{2}{h} \, D \log P (z , \theta) \, \mu \cdot H 
  + m e^\lambda \, P (z , \theta)^{\! \frac{1}{h}} 
  \left( \sum_{i = 1}^{r} E_{\alpha_i} + E_{\alpha_0} \!\! \right) . 
  \label{eq:gauge_connection}
\end{equation}
The rescalings introduced below are chosen so that $x$, $E$, and $\tilde{\theta}$ remain finite in the limit $\lambda \to \infty$, $\bar{z} \to 0$, and $z \to 0$. 
After taking this limit, we apply the inverse gauge transformation and recover a simpler Toda-type Lax operator, which will serve as the starting point for the subsequent diagonalization and WKB analysis. 
We rescale $z , \bar{z} ,\, s , \bar{s} ,\, \theta$ and $\bar{\theta}$ as 
\begin{equation}
  \begin{aligned}
    x &= \left( m e^\lambda \right)^{\! \frac{2}{M + 1}} z , & 
    E &= s^{\frac{h M}{2}} \left( m e^\lambda \right)^{\! \frac{h M}{M + 1}} , &
    \tilde{\theta} &= \theta \left( m e^\lambda \right)^{\! \frac{1}{M + 1}} ,  \\
    \bar{x} &= \left( m e^{- \lambda} \right)^{\! \frac{2}{M + 1}} \bar{z} , & 
    \bar{E} &= \bar{s}^{\frac{h M}{2}} \left( m e^{- \lambda} \right)^{\! \frac{h M}{M + 1}} , & 
    \tilde{\bar{\theta}} &= \bar{\theta} \left( m e^{- \lambda} \right)^{\! \frac{1}{M + 1}} , 
  \end{aligned}
  \label{eq:rescaling}
\end{equation}
and take the light-cone limit $\bar{z} \to 0$ with $\lambda \to \infty$. 
Then we take the conformal limit $z \to 0$ with $x ,\, E$ kept finite. 
In this limit, the differential equation for the super-Lax operator $U \mathcal{L}_{\mathrm{F}} U^{- 1} = D - A^U_\theta$ reduces to the equation $\mathcal{L}'_{\mathrm{F}} \varPsi = 0$, where 
\begin{equation}
  \mathcal{L}'_{\mathrm{F}} 
  = \mathcal{D} - \frac{\varLambda (\tilde{\theta}) \cdot H}{x} 
  + \frac{2}{h} \, \mathcal{D} \log P (x , \tilde{\theta} , E) \, \mu \cdot H 
  - P (x , \tilde{\theta} , E)^{\! \frac{1}{h}} \left( \sum_{i = 1}^{r} E_{\alpha_i} + E_{\alpha_0} \!\! \right) 
  \label{eq:Lax_reduction}
\end{equation}
with $\mathcal{D} = \partial_{\tilde{\theta}} + \tilde{\theta} \, \partial_x , \; 
\varLambda (\tilde{\theta}) = \xi + \tilde{\theta} \, l , \; 
P (x , \tilde{\theta} , E) = p_0 (x , E) + \tilde{\theta} \, p_1 (x , E)$. 
After the rescaling \eqref{eq:rescaling}, the light-cone limit, and the conformal limit, the bosonic part of the superspace potential takes the form $p_0 (x , E) = x^{\frac{h M}{2}} - E$. 
We further rescale the variables as
\begin{equation}
  x \to \epsilon^{- \frac{1}{M + 1}} x , \qquad
  E \to \epsilon^{- \frac{h M}{2 (M + 1)}} E , \qquad
  \tilde{\theta} \to \epsilon^{- \frac{1}{2 (M + 1)}} \tilde{\theta} ,
\end{equation}
and simultaneously
\begin{equation}
  \xi \to \epsilon^{- \frac{1}{2 (M + 1)}} \xi , \qquad
  p_1 (x) \to \epsilon^{- \frac{h M - 1}{2 (M + 1)}} p_1 (x) .
\end{equation}
Under this rescaling, the bosonic part of the potential is brought to the form 
$p_0 (x , E)=x^{\frac{h M}{2}} - E$ up to an overall factor, and we may further fix $E = 1$. 
Accordingly, the super-Lax operator takes the form 
\begin{equation}
  \mathcal{L}'_{\mathrm{F}}
  = \epsilon^{\frac{1}{2 (M + 1)}} \, \mathcal{D} 
  - \frac{\epsilon^{\frac{1}{2 (M + 1)}} \, \varLambda (\tilde{\theta}) \cdot H}{x} 
  + \frac{2}{h} \, \epsilon^{\frac{1}{2 (M + 1)}} \, \mathcal{D} \log P (x , \tilde{\theta}) \, \mu \cdot H 
  - \epsilon^{- \frac{M}{2 (M + 1)}} P (x , \tilde{\theta})^{\frac{1}{h}} 
  \left( \sum_{i = 1}^r E_{\alpha_i} + E_{\alpha_0} \right) . 
\end{equation}
Multiplying the linear problem by an overall power of $\epsilon$, one can rewrite it in the equivalent form 
\begin{equation}
  \mathcal{L}'_{\mathrm{F}} 
  = \epsilon^{\frac{1}{2}} \, \mathcal{D} 
  - \frac{\epsilon^{\frac{1}{2}} \varLambda (\tilde{\theta}) \cdot H}{x} 
  + \frac{2}{h} \, \epsilon^{\frac{1}{2}} \, \mathcal{D} \log P (x , \tilde{\theta}) \, \mu \cdot H 
  - P (x , \tilde{\theta})^{\! \frac{1}{h}} 
  \left( \sum_{i = 1}^{r} E_{\alpha_i} + E_{\alpha_0} \!\! \right) , 
\end{equation}
where $\epsilon$ plays the role of the Planck constant in the WKB expansion. 
\par
Applying inverse gauge transformation with the gauge parameter $\exp \left[ - \frac{2}{h} \log P (x , \tilde{\theta}) \, \mu \cdot H \right]$, one can obtain the Lax operator 
\begin{equation}
  \mathcal{L}_{\mathrm{F}} = \epsilon^{\frac{1}{2}} \, \mathcal{D} 
  - \frac{\epsilon^{\frac{1}{2}} \varLambda (\tilde{\theta}) \cdot H}{x} 
  - \left[ \sum_{i = 1}^{r} E_{\alpha_i} + P (x , \tilde{\theta}) \, E_{\alpha_0} \! \right] , 
  \label{eq:super-Lax_reduction}
\end{equation}
which leads to the bosonic Lax operator of the form 
\begin{align}
  \mathcal{L}_{\mathrm{B}} =\epsilon \, \partial_x 
  &- \frac{\epsilon \, l \cdot H}{x} 
  - \epsilon^{\frac{1}{2}} \, p_1 (x) \, E_{{\alpha_0}} \notag \\
  &- \left\{ \! \left[ \sum_{i = 1}^{r} E_{\alpha_i} + p_0 (x) \, E_{\alpha_0} \! \right] 
  - \frac{\epsilon^{\frac{1}{2}} \, \xi \cdot H}{x} \! \right\} \! 
  \left\{ \! \left[ \sum_{i = 1}^{r} E_{\alpha_i} + p_0 (x) \, E_{\alpha_0} \! \right] 
  + \frac{\epsilon^{\frac{1}{2}} \, \xi \cdot H}{x} \! \right\} . 
  \label{eq:bosonic-Lax_reduction}
\end{align}
In the case where the Cartan subalgebra is one-dimensional, the bosonic Lax operator reduces to 
\begin{equation}
  \mathcal{L}_{\mathrm{B}} = \epsilon \, \partial_x 
  - \frac{\epsilon \, l \cdot H}{x} 
  - \epsilon^{\frac{1}{2}} \, p_1 (x) \, E_{\alpha_0} 
  - [E_{\alpha_1} + p_0 (x) \, E_{\alpha_0}]^2 
  + \left[ \frac{\epsilon^{\frac{1}{2}} \, \xi \cdot H}{x} , E_{\alpha_1} \right] 
  + p_0 (x) \left[ \frac{\epsilon^{\frac{1}{2}} \, \xi \cdot H}{x} , E_{\alpha_0} \right] . 
  \label{eq:1d_bosonic-Lax}
\end{equation}
In this case, $\xi \cdot H$ is controlled by a single Grassmann-odd parameter, and its nilpotency simplifies the quadratic expression in Eq.~\eqref{eq:bosonic-Lax_reduction}. 
Accordingly, we consider the bosonic linear problem 
\begin{equation}
  \mathcal{L}_{\mathrm{B}} \varPsi_0 (x) = [\epsilon \, \partial_x + A (x)] \varPsi_0 (x) = 0 , 
  \label{eq:bosonic_linear_problem}
\end{equation}
with
\begin{equation}
  A (x) \coloneqq - \frac{\epsilon \, l \cdot H}{x} 
  - \epsilon^{\frac{1}{2}} \, p_1 (x) \, E_{\alpha_0} 
  - [E_{\alpha_1} + p_0 (x) \, E_{\alpha_0}]^2 
  + \left[ \frac{\epsilon^{\frac{1}{2}} \, \xi \cdot H}{x} , E_{\alpha_1} \right] 
  + p_0 (x) \left[ \frac{\epsilon^{\frac{1}{2}} \, \xi \cdot H}{x} , E_{\alpha_0} \right] . 
  \label{eq:connection_1d}
\end{equation}
The reduced form \eqref{eq:1d_bosonic-Lax} will be the starting point for the $C (2)^{(2)}$ analysis in the next section. 

%% file: C22_linear.tex
\section{Diagonalization of \texorpdfstring{$\bm{C (2)^{(2)}}$}{$C (2)^{(2)}$}-type linear problem}
\label{sec:linear_problem}
In this section, we specialize the general construction of Section~\ref{sec:super-Toda} to the $C (2)^{(2)} = \mathfrak{osp} (2 | 2)^{(2)}$ case and study the associated linear problem in detail. 
We first introduce the $C (2)^{(2)}$ linear problem and its three-dimensional representation, and then diagonalize the connection by a sequence of gauge transformations following \cite{ito2023a,ide2026}. 
This leads to the WKB periods that will later be compared with the local integrals of motion on the SCFT side. 

\subsection{The diagonalization approach to the linear problem}
\label{subsec:diagonalization}
We now diagonalize the bosonic linear problem \eqref{eq:bosonic_linear_problem} by a sequence of gauge transformations. 
The resulting diagonal connection determines the WKB solutions and the associated WKB periods. 
\par
To this end, we introduce a gauge transformation of the form 
\begin{equation}
  \mathbf{Gau}_T [A (x)] 
  = T^{- 1} (x) \, A (x) \, T (x) + \epsilon \, T^{- 1} (x) \, \partial_x T (x) , 
\end{equation}
where 
\begin{equation}
  T (x) \coloneqq T_d (x) \, T_{d - 1} (x) \cdots T_1 (x) , 
\end{equation}
and each elementary matrix $T_i (x)$ differs from the identity only in the $i$-th row: 
\begin{equation}
  T_i (x)_{a b} \coloneqq 
  \begin{cases}
    1 & (a = b) , \\
    g_{i ,\, b} (x) & (a = i ,\; b \neq i ,\; 1 \leq b \leq d) , \\
    0 & (\text{otherwise}) . 
  \end{cases}
\end{equation}
At the $i$-th step, the functions $g_{i ,\, b} (x)$ are fixed so that the transformed connection $A' (x) = \mathbf{Gau}_{T_i} [A (x)]$ satisfies 
\begin{equation}
  A'_{i j} = 0 , \quad 1 \leq j \leq d , \quad j \neq i . 
  \label{eq:Riccati_eq}
\end{equation}
In other words, we eliminate the off-diagonal entries in the $i$-th row order by order in $\epsilon^{\frac{1}{2}}$. 
These conditions give a coupled set of Riccati-type constraints for the gauge parameters. 
Iterating this procedure row by row, we obtain a diagonal connection 
\begin{equation}
  A_{\mathrm{diag}} (x) = \mathbf{Gau}_{T_1} \circ \mathbf{Gau}_{T_2} \circ \cdots \circ \mathbf{Gau}_{T_d} [A (x)] , 
\end{equation}
whose diagonal entries are written as 
\begin{equation}
  A_{\mathrm{diag}} (x) = \operatorname{diag} \left\{ f_1 (x) , f_2 (x) , \dots , f_d (x) \right\} . 
\end{equation}
\par
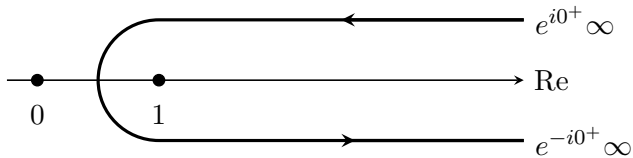
\begin{figure}[H]
  \centering
  \begin{tikzpicture}[>=stealth,scale=0.8]
    \begin{feynhand}
      \draw[semithick,->] (-0.5,0) -- (8,0) node [right] {$\Re$};
      \vertex[dot] (a) at (0,0) {};
      \vertex[dot] (b) at (2,0) {};
      \node at (0,-0.25) [below] {0};
      \node at (2,-0.25) [below] {1};
      \draw[very thick] (8,1) node [right] {$e^{i 0^+} \infty$} -- (2,1) arc (90:270:1) -- (8,-1) node [right] {$e^{- i 0^+} \infty$};
      \draw[very thick,->] (8, 1) -- (5, 1);
      \draw[very thick,-<] (8,-1) -- (5,-1);
    \end{feynhand}
  \end{tikzpicture}
  \caption{The Pochhammer contour $\mathcal{C}$. }
  \label{fig:Pochhammer}
\end{figure}

The transformed linear problem then reduces to a set of decoupled first-order equations, 
\begin{equation}
  [\epsilon \, \partial_x + f_i (x)] \varPsi_i (x) = 0 \quad 
  (i = 1 , \dots , d) , 
\end{equation}
whose solutions take the WKB form 
\begin{equation}
  \varPsi_i (x) \sim \exp \left[ - \frac{1}{\epsilon} \int^x f_i \left( x' \right) \, dx' \right] . 
\end{equation}
The WKB periods associated with the $i$-th diagonal component are defined by the contour integrals over a cycle $\mathcal{C}$ on the complex plane, 
\begin{equation}
  \oint_{\mathcal{C}} f_i (z) \, dz = \sum_{k = 0}^{\infty} \epsilon^k Q_k^{(i)} . 
  \label{eq:WKB_period}
\end{equation}
In this paper, we take $\mathcal{C}$ to be the Pochhammer contour shown in Fig.\,\ref{fig:Pochhammer}, following \cite{babenko2017}. 
We refer to $Q_k^{(i)}$ as the $k$-th WKB period. 

\subsection[The diagonalization of $C (2)^{(2)}$-type linear problem]
{The diagonalization of \texorpdfstring{$\bm{C (2)^{(2)}}$}{$C (2)^{(2)}$}-type linear problem}
\label{subsec:diag_C(2)^(2)}
\begin{figure}[H]
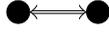

  \centering 
  \begin{dynkinDiagram}[
      labels*={
        {\alpha_0},
        {\alpha_1}
      },
      edge length={1.5cm},
      root radius={0.15cm},
      text style/.style={scale=0.85},
      line width=5mm,
      arrow width=1mm,
      extended,
      mark=*,
      affine mark=*
    ]{A}{1}
  \end{dynkinDiagram}
  \caption{Dynkin diagram of the twisted affine Lie superalgebra $C (2)^{(2)}$, 
  where a black dot represents the odd simple root with non-zero length. }
  \label{fig:C(2)^(2)}
\end{figure}

We now specialize the general construction to the case of $C (2)^{(2)} = \mathfrak{osp} (2 | 2)^{(2)}$, and apply the diagonalization procedure of Subsection~\ref{subsec:diagonalization} to the corresponding linear problem. 
There are two simple fermionic roots $\beta_1 = \delta_1 - e_1$ and $\beta_2 = e_1 + \delta_1$. 
The twisted affine Lie algebra is defined by the $\mathbb{Z}_2$-automorphism $e_1 \to - e_1$, under which these two roots are exchanged. 
The resulting simple root is $\alpha_1 = \frac{1}{2} (\beta_1 + \beta_2) = \delta_1$, 
while the extended root (an odd root) is the lowest weight $\alpha_0 = - \frac{1}{2} (\beta_1 + \beta_2) = - \delta_1$. 
The Dynkin diagram of $C (2)^{(2)}$ is shown in Fig.\,\ref{fig:C(2)^(2)}. 
The generators of $\mathfrak{osp} (2 | 2)^{(2)}$ are given by 
\begin{equation}
  \begin{aligned}
    E_{\alpha_1} &= \frac{1}{\sqrt{2}} (E_{\beta_1} + E_{\beta_2}) , & 
    E_{\alpha_0} &= \frac{1}{\sqrt{2}} (E_{- \beta_1} - E_{- \beta_2}) , \\
    E_{- \alpha_1} &= \frac{1}{\sqrt{2}} (E_{- \beta_1} + E_{- \beta_2}) , & 
    E_{- \alpha_0} &= \frac{1}{\sqrt{2}} (E_{\beta_1} - E_{\beta_2}) . 
  \end{aligned}
  \label{eq:C(2)^(2)_root}
\end{equation}
Then the modified affine Toda field equation \eqref{eq:mod_affine_Toda_eq} becomes 
\begin{equation}
  D \bar{D} \varPhi 
  + \frac{m^2}{\beta} \, \alpha_1 \, e^{\beta \alpha_1 \cdot \varPhi} 
  + \frac{m^2}{\beta} \, P (z , \theta) \, \bar{P} \left( \bar{z} , \bar{\theta} \, \right) \, 
  \alpha_0 \exp (\beta \, \alpha_0 \cdot \varPhi) = 0 , 
  \label{eq:ssG}
\end{equation}
which is nothing but the (modified) $\mathcal{N} = 1$ supersymmetric sinh-Gordon equation \cite{kulish2005}. 
The related linear problem and its bosonic reduction can be written down using the matrix representation of the generators. 
We work in the three-dimensional representation throughout. 
The generators are 
\begin{equation}
  \begin{aligned}
    E_{\alpha_1} &= E_{1 ,\, 2} + E_{2 ,\, 3} , \quad E_{- \alpha_1} = E_{3 ,\, 2} - E_{2 ,\, 1} , \\
    E_{\alpha_0} &= - E_{2 ,\, 1} - E_{3 ,\, 2} , \quad E_{- \alpha_0} = E_{1 ,\, 2} - E_{2 ,\, 3} , \\
    \alpha_1 \cdot H &= - \alpha_0 \cdot H = \{ E_{\alpha_1} , E_{- \alpha_1} \} 
    = \{ E_{\alpha_0} , E_{- \alpha_0} \} = E_{3 ,\, 3} - E_{1 ,\, 1} . 
  \end{aligned}
  \label{eq:3d-rep}
\end{equation}
The super-Lax operator is given by 
\begin{equation}
  \mathcal{L}_{\mathrm{F}} = \sqrt{\epsilon} \, \mathcal{D} + 
  \begin{pmatrix}
    \frac{\sqrt{\epsilon} \, \varLambda}{x} & - 1 & 0 \\
    P & 0 & - 1 \\
    0 & P & - \frac{\sqrt{\epsilon} \, \varLambda}{x} 
  \end{pmatrix}
  . 
  \label{eq:3d-super-Lax}
\end{equation}
The bosonic Lax operator is given by 
\begin{equation}
  \mathcal{L}_{\mathrm{B}} (x) = \epsilon \, \partial_x + A (x) 
  = \epsilon \, \partial_x + 
  \begin{pmatrix}
    \frac{\epsilon \, l}{x} + p_0 (x) & - \frac{\sqrt{\epsilon} \, \xi}{x} & - 1 \\
    \sqrt{\epsilon} \, p_1 (x) - \frac{\sqrt{\epsilon} \, \xi}{x} \, p_0 (x) & 2 p_0 (x) & - \frac{\sqrt{\epsilon} \, \xi}{x} \\
    - p_0 (x)^2 & - \sqrt{\epsilon} \, p_1 (x) - \frac{\sqrt{\epsilon} \, \xi}{x} \, p_0 (x) & - \frac{\epsilon \, l}{x} + p_0 (x) 
  \end{pmatrix}
  . 
  \label{eq:3d-Lax}
\end{equation}
\par
For the three-dimensional representation of $C (2)^{(2)}$, we regard the connection as a supermatrix of block type 
\begin{equation}
  \left(
    \begin{array}{c|c|c}
      \mathrm{B} \mathrm{B} & \mathrm{B} \mathrm{F} & \mathrm{B} \mathrm{B} \\
      \hline 
      \quad & \quad & \quad \\[-4.5mm]
      \mathrm{F} \mathrm{B} & \mathrm{F} \mathrm{F} & \mathrm{F} \mathrm{B} \\
      \hline 
      \quad & \quad & \quad \\[-4.5mm]
      \mathrm{B} \mathrm{B} & \mathrm{B} \mathrm{F} & \mathrm{B} \mathrm{B} 
    \end{array}
  \right) , 
\end{equation}
where $\mathrm{B}$ and $\mathrm{F}$ denote even and odd components, respectively. 
Accordingly, the supertrace is given by 
\begin{equation}
  \operatorname{Str} A (x) = A_{11} (x) - A_{22} (x) + A_{33} (x) . 
\end{equation}
Since the connection \eqref{eq:3d-Lax} is supertraceless, the first diagonal component is determined once the second and third diagonal components are known. 
Therefore, it is sufficient to diagonalize the third and second rows explicitly. 
\par
The Grassmann parity of the gauge parameters is assigned according to the above block structure: 
\begin{equation}
  \begin{aligned}
    |g_{3 ,\, 1} (x)| &= 0 , & 
    |g_{2 ,\, 1} (x)| &= |g_{2 ,\, 3} (x)| = |g_{3 ,\, 2} (x)| = 1 , \\
    |f_i (x)| &= 0 & (i &= 1 ,2 , 3) . 
  \end{aligned}
\end{equation}

\subsubsection[Branch structure and the vanishing of $p_1 (x)$]
{Branch structure and the vanishing of \texorpdfstring{$\bm{p_1 (x)}$}{$p_1 (x)$}}
\label{subsubsec:branch}
Before proceeding to the higher-order diagonalization, let us first examine the leading-order structure of the third-row constraints. 
This analysis reveals that the classical solution admits two branches, and that the behavior of the fermionic quantity $p_1 (x)$ depends on the branch. 
These branch-dependent features will play an important role in the subsequent diagonalization.
\par
To diagonalize the third row, we introduce the elementary gauge transformation $T_3 (x)$, for which the transformed connection takes the form 
\begin{equation}
  \mathbf{Gau}_{T_3} [A (x)] = 
  \begin{pmatrix}
    \frac{\epsilon \, l}{x} + p_0 - g_{3 ,\, 1} & - \frac{\sqrt{\epsilon} \, \xi}{x} - g_{3 ,\, 2} & - 1 \\
    - \frac{\sqrt{\epsilon} \, \xi}{x} (g_{3 ,\, 1} + p_0) + \sqrt{\epsilon} \, p_1 
    & - \frac{\sqrt{\epsilon} \, \xi}{x} \, g_{3 ,\, 2} + 2 p_0 & - \frac{\sqrt{\epsilon} \, \xi}{x} \\
    r_{31} (x) & r_{32} (x) & f_3 (x) 
  \end{pmatrix}
   , 
  \label{eq:GauT3}
\end{equation}
where
\begin{align}
  r_{31} (x) &= \sqrt{\epsilon} \, p_1 g_{3 ,\, 2} - p_0^2 - \frac{2 \epsilon \, l}{x} \, g_{3 ,\, 1} + g_{3 ,\, 1}^2 
  - \frac{\sqrt{\epsilon} \, \xi}{x} (p_0 + g_{3 ,\, 1}) g_{3 ,\, 2} + \epsilon \, g'_{3 ,\, 1} , 
  \label{eq:Gau31} \\
  r_{32} (x)  &= 
  \frac{\sqrt{\epsilon} \, \xi}{x} (g_{3 ,\, 1} - p_0) + \sqrt{\epsilon} \, p_1 
  - \frac{\epsilon \, l}{x} \, g_{3 ,\, 2} + (g_{3 ,\, 1} - p_0) g_{3 ,\, 2} + \epsilon \, g'_{3 ,\, 2} , 
  \label{eq:Gau32} \\
  f_3 (x) &= - \frac{\epsilon \, l}{x} - \frac{\sqrt{\epsilon} \, \xi}{x} \, g_{3 ,\, 2} + p_0 + g_{3 ,\, 1} . 
  \label{eq:f3}
\end{align}
We expand the gauge parameters in a WKB form compatible with the $\epsilon$-scaling and the Grassmann parity assignment: 
\begin{equation}
  g_{3 ,\, 1} (x) = \sum_{k = 0}^{\infty} \epsilon^k \, S_{3 ,\, 1}^{(k)} (x) , \quad 
  g_{3 ,\, 2} (x) = \sum_{k = 0}^{\infty} \epsilon^{k + \frac{1}{2}} \, S_{3 ,\, 2}^{(k)} (x) , 
  \label{eq:WKB_g3}
\end{equation}
Here the half-integer shift in the expansion of $g_{3 ,\, 2} (x)$ reflects the fact that $g_{3 ,\, 2} (x)$ is Grassmann-odd and appears in the connection together with fermionic quantities of order $\epsilon^{\frac{1}{2}}$. 
Substituting these expansions into the Riccati-type constraints $r_{31} (x) = 0$ and $r_{32} (x) = 0$, we solve them order by order in $\epsilon^{\frac{1}{2}}$. 
At the leading order $\epsilon^0$, the $(3 , 1)$-constraint reduces to 
\begin{equation}
  S_{3 ,\, 1}^{(0)} (x)^2 - p_0 (x)^2 = 0 , 
  \label{eq:S310}
\end{equation}
so that there are two classical branches
\begin{equation}
  S_{3 ,\, 1}^{(0)} (x) = \pm p_0 (x) . 
  \label{eq:S310_branches}
\end{equation}
Once a branch is chosen, the $(3 , 2)$-constraint at the leading order $\epsilon^{\frac{1}{2}}$ determines the leading coefficient $S_{3 ,\, 2}^{(0)}$. 
Indeed, keeping only the $\epsilon^{\frac{1}{2}}$-terms in $r_{32} (x) = 0$, we obtain 
\begin{equation}
  \left[ S_{3 ,\, 1}^{(0)} (x) - p_0 (x) \right] S_{3 ,\, 2}^{(0)} (x) 
  + \frac{\xi}{x} \left[ S_{3 ,\, 1}^{(0)} (x) - p_0 (x) \right] + p_1 (x) = 0 . 
  \label{eq:S320}
\end{equation}
In particular, in the minus branch $S_{3 ,\, 1}^{(0)} (x) = - p_0 (x)$, this gives 
\begin{equation}
  S_{3 ,\, 2}^{(0)} (x) = - \frac{\xi}{x} + \frac{p_1 (x)}{2 p_0 (x)} . 
  \label{eq:S320_minus}
\end{equation}
In the plus branch $S_{3 ,\, 1}^{(0)} (x) = + p_0 (x)$, one instead finds 
\begin{equation}
  p_1 (x) = 0 . 
  \label{eq:plus_branch_p1_zero}
\end{equation}
At this order, however, $S_{3 ,\, 2}^{(0)} (x)$ remains undetermined. 
Finally, the classical part of the diagonal component $f_3 (x)$ is read off from Eq.~\eqref{eq:f3}, so that at the leading order we have 
\begin{equation}
  f_3^{(0)} (x) = p_0 (x) + S_{3 ,\, 1}^{(0)} (x) = 
  \begin{cases}
    0 & \text{(minus branch)} , \\
    2 p_0 (x) & \text{(plus branch)} . 
  \end{cases}
  \label{eq:f30}
\end{equation}
\par
At this stage, the status of $p_1 (x)$ remains branch-dependent. 
To determine whether the fermionic quantity $p_1 (x)$ survives in the minus branch, we must also examine the leading-order consistency of the second-row constraint. 
We therefore next study the second-row diagonalization in the minus branch and compare the resulting condition with the plus-branch result. 
\par
After the second  gauge transformation $T_2$, the connection becomes 
\begin{equation}
  \mathbf{Gau}_{T_2 T_3} [A (x)] = 
  \begin{pmatrix}
    \frac{\epsilon \, l}{x} - \frac{\sqrt{\epsilon} \, \xi}{x} \, g_{2 ,\, 1} 
    - g_{3 ,\, 2} g_{2 ,\, 1} + p_0 - g_{3 ,\, 1} & 
    - \frac{\sqrt{\epsilon} \, \xi}{x} - g_{3 ,\, 2} & 
    - \frac{\sqrt{\epsilon} \, \xi}{x} \, g_{2 ,\, 3} - g_{3 ,\, 2} g_{2 ,\, 3} - 1 \\
    r_{21} (x) & f_2 (x) & r_{23} (x) \\
    0 & 0 & f_3 (x) 
  \end{pmatrix}
  , 
  \label{eq:GauT2}
\end{equation}
where
\begin{align}
  r_{21} (x) 
  &= - \frac{\sqrt{\epsilon} \, \xi}{x} (g_{3 ,\, 2} g_{2 ,\, 1} + g_{3 ,\, 1} + p_0) + \sqrt{\epsilon} \, p_1 
  - \frac{\epsilon \, l}{x} \, g_{2 ,\, 1} + p_0 g_{2 ,\, 1} + g_{2 ,\, 1} g_{3 ,\, 1} + \epsilon \, g'_{2 ,\, 1} , 
  \label{eq:Gau21} \\
  f_2 (x) &= 2 p_0 - g_{3 ,\, 2} g_{2 ,\, 1} - \frac{\sqrt{\epsilon} \, \xi}{x} (g_{2 ,\, 1} - g_{3 ,\, 2}) , 
  \label{eq:f2} \\
  r_{23} (x) 
  &= - \frac{\sqrt{\epsilon} \, \xi}{x} (g_{2 ,\, 1} g_{2 ,\, 3} + g_{3 ,\, 2} g_{2 ,\, 3} + 1) 
  - g_{3 ,\, 2} g_{2 ,\, 1} g_{2 ,\, 3} + g_{2 ,\, 1} + f_3 \, g_{2 ,\, 3} - 2 p_0 g_{2 ,\, 3} 
  - \epsilon \, g'_{2 ,\, 3} . 
  \label{eq:Gau23}
\end{align}
Since the diagonal entry $f_2 (x)$ does not contain $g_{2 ,\, 3} (x)$, the determination of $f_2 (x)$ at a given order only requires solving the constraint 
$r_{21} (x) = 0$ for $g_{2 ,\, 1} (x)$. 
The equation $r_{23} (x) = 0$, which fixes $g_{2 ,\, 3} (x)$, can be postponed because it is irrelevant for the computation of $f_2 (x)$ at this stage. 
\par
We now analyze $r_{21} (x) = 0$ in the minus branch of the third-row diagonalization. 
Recall that the third-row gauge parameters were expanded as in Eq.~\eqref{eq:WKB_g3}, and that the minus branch is characterized by Eqs.\,\eqref{eq:S310_branches} and \eqref{eq:S320_minus}. 
For $g_{2 ,\, 1} (x)$, we likewise adopt a fermionic WKB expansion compatible with the $\epsilon$-scaling, 
\begin{equation}
  g_{2 ,\, 1} (x) = \sum_{k = 0}^{\infty} \epsilon^{k + \frac{1}{2}} \, S_{2 ,\, 1}^{(k)} (x) . 
\end{equation}
The half-integer shift reflects the fact that $g_{2 ,\, 1} (x)$ is Grassmann-odd and therefore starts at the same semiclassical order as the fermionic quantities $p_1 (x)$ and $\xi$. 
At the leading order $\epsilon^{\frac{1}{2}}$, the $(2 , 1)$-constraint reduces to 
\begin{equation}
  S_{2 ,\, 1}^{(0)} (x) \left[ S_{3 ,\, 1}^{(0)} (x) + p_0 (x) \right] 
  - \frac{\xi}{x} \left[ S_{3 ,\, 1}^{(0)} (x) + p_0 (x) \right] + p_1 (x) = 0 , 
  \label{eq:S210}
\end{equation}
In the minus branch we have $S_{3 ,\, 1}^{(0)} (x) = - p_0 (x)$, and therefore 
\begin{equation}
  p_1 (x) = 0 . 
\end{equation}
Hence, for $C(2)^{(2)}$ we find $p_1 (x) = 0$ in both the minus and the plus branches. 
\par
Moreover, the computation of the third row in the plus branch agrees with that of the first row in the minus branch. 
Since the connection is supertraceless, the first diagonal component is determined once the second and third diagonal components are known. 
Therefore, it is sufficient in what follows to work in the minus branch and analyze only the third and second rows. 

\subsubsection{Diagonalization of the third row}
\label{subsubsec:3rd_diag}
We now proceed to the higher-order diagonalization of the third row. 
The corresponding WKB coefficients can be determined recursively from the super-Riccati system. 
We expand the gauge parameters and the diagonal entry as 
\begin{equation}
  g_{3 ,\, 1} (x) = \sum_{k = 0}^{\infty} \epsilon^k \, S_{3 ,\, 1}^{(k)} (x) , \quad 
  g_{3 ,\, 2} (x) = \sum_{k = 0}^{\infty} \epsilon^{k + \frac{1}{2}} \, S_{3 ,\, 2}^{(k)} (x) , \quad 
  f_3 (x) = \sum_{k = 0}^{\infty} \epsilon^k \, f_3^{(k)} (x) . 
\end{equation}
Accordingly, we also expand 
\begin{equation}
  r_{31} (x) = \sum_{k = 0}^{\infty} \epsilon^k \, r_{31}^{(k)} (x) , \quad 
  r_{32} (x) = \sum_{k = 0}^{\infty} \epsilon^{k + \frac{1}{2}} \, r_{32}^{(k)} (x) . 
\end{equation}
Substituting these expansions into Eqs.\,\eqref{eq:Gau31}--\eqref{eq:f3} and collecting terms order by order in $\epsilon^{\frac{1}{2}}$, we obtain a hierarchy of algebraic-differential equations. 
At the leading orders, we obtain 
\begin{align}
  r_{31}^{(0)} (x) &= - p_0 (x)^2 + {S_{3 ,\, 1}^{(0)}} (x)^2 = 0 , \\
  r_{32}^{(0)} (x) &= - \frac{\xi}{x} \, p_0 (x) + \frac{\xi}{x} \, S_{3 ,\, 1}^{(0)} (x) 
  - p_0 S_{32}^{(0)} + S_{3 ,\, 1}^{(0)} (x) S_{3 ,\, 2}^{(0)} (x) = 0 , \\
  f_3^{(0)} (x) &= p_0 (x) + S_{3 ,\, 1}^{(0)} (x) , 
\end{align}
while the next orders give 
\begin{align}
  r_{31}^{(1)} (x) &= - \frac{\xi}{x} \, p_0 S_{3 ,\, 2}^{(0)} 
  - \frac{2 l}{x} \, S_{3 ,\, 1}^{(0)} 
  - \frac{\xi}{x} \, S_{31}^{(0)} S_{3 ,\, 2}^{(0)} 
  + 2 S_{3 ,\, 1}^{(0)} S_{3 ,\, 1}^{(1)} + S_{3 ,\, 1}^{(0) \prime} = 0 , \\
  r_{32}^{(1)} (x) &= - \frac{\xi}{x} \, S_{3 ,\, 1}^{(1)} 
  - \frac{l}{x} \, S_{3 ,\, 2}^{(0)} + S_{3 ,\, 1}^{(1)} S_{3 ,\, 2}^{(0)} 
  - p_0 S_{3 ,\, 2}^{(1)} + S_{3 ,\, 1}^{(0)} S_{3 ,\, 2}^{(1)} 
  + S_{3 ,\, 2}^{(0) \prime} = 0 , \\
  f_3^{(1)} (x) &= - \frac{l}{x} - \frac{\xi}{x} \, S_{3 ,\, 2}^{(0)} + S_{3 ,\, 1}^{(1)} .
\end{align}
\par
To organize the higher-order equations, it is convenient to introduce the vectors 
\begin{equation}
  \bm{R}^{(k)} (x) \coloneqq 
  \begin{pmatrix}
    r_{31}^{(k)} (x) \\[1mm]
    r_{32}^{(k)} (x) 
  \end{pmatrix}
  , \quad 
  \bm{S}^{(k)} (x) \coloneqq
  \begin{pmatrix}
    S_{3 ,\, 1}^{(k)} (x) \\[1mm]
    S_{3 ,\, 2}^{(k)} (x) 
  \end{pmatrix}
  . 
\end{equation}
At each order, the unknown coefficients enter linearly, while all nonlinear contributions are determined by lower-order data already fixed in the previous steps, as in the recursive diagonalization method proposed in \cite{ide2026}. 
Therefore, the super-Riccati system at order $\epsilon^k$ can be written in the form 
\begin{equation}
  \bm{R}^{(k)} (x) = - B (x) \, \bm{S}^{(k)} (x) + \bm{J}^{(k)} (x) , 
  \label{eq:Rk_structure}
\end{equation}
where $\bm{J}^{(k)} (x)$ depends only on the lower-order coefficients $\bm{S}^{(j)} (x)$ with $j < k$, which have already been determined recursively. 
The matrix $B (x)$ is the Jacobian of the leading-order constraints, 
\begin{equation}
  B_{a b}(x) 
  \coloneqq - \frac{\partial r_{3 a}^{(0)} (x)}{\partial S_{3 ,\, b}^{(0)} (x)} 
  \quad (a , b = 1 , 2) , 
\end{equation}
where the derivative with respect to the Grassmann-odd variable $S_{3 ,\, 2}^{(0)} (x)$ is taken to be the left derivative. 
In the present case, it is explicitly given by 
\begin{equation}
  B (x) = 
  \begin{pmatrix}
    - 2 S_{3 ,\, 1}^{(0)} (x) & 0 \\
    - \frac{\xi}{x} + S_{3 ,\, 2}^{(0)} (x) & - 2 S_{3 ,\, 1}^{(0)} (x) 
  \end{pmatrix}.
  \label{eq:Bmatrix}
\end{equation}
Imposing the constraints $\bm{R}^{(k)} (x) = 0$, we obtain the recursive solution 
\begin{equation}
  \bm{S}^{(k)} (x) = B^{- 1} (x) \, \bm{J}^{(k)} (x) . 
  \label{eq:Sk_recursive}
\end{equation}
Thus all higher-order coefficients are determined recursively from the lower-order data. 
\par
With $B (x)$ at hand, the higher-order coefficients are determined recursively, and the diagonal components $f_3^{(k)} (x)$ can therefore be obtained order by order. 
The explicit expressions for $f_3^{(k)} (x)$ up to $k = 6$ are 
\begin{align}
  f_3^{(0)} (x) &= 0 , \\
  f_3^{(1)} (x) &= - \frac{p'_0 (x)}{2 p_0 (x)} 
  = - \frac{1}{2} \frac{d}{dx} \log p_0 (x) = d(*) , \\
  f_3^{(2)} (x) &= - \frac{l (l + 1)}{2 x^2 p_0 (x)} - \frac{p''_0 (x)}{16 p_0 (x)^2} + d(*) , \\
  f_3^{(3)} (x) &= d(*) , \\
  f_3^{(4)} (x) &= \frac{l \left( 4 (l + 2) l^2 + l - 3 \right)}{32 x^4 p_0 (x)^3} 
  + \frac{5 l (l + 1) p''_0 (x)}{64 x^2 p_0 (x)^4} 
  - \frac{3 p_0^{(4)} (x)}{1024 p_0 (x)^4} 
  + \frac{5 p''_0 (x)^2}{256 p_0 (x)^5} 
  + d(*) , \\
  f_3^{(5)} (x) &= d (*) , \\
  f_3^{(6)} (x) &= - \frac{l (l + 1) (16 l (l + 1) (6 l (l + 1) - 13) + 135)}{1536 x^6 p_0 (x)^5} \notag \\
  &\quad - \frac{35 l (l + 1) (8 l (l + 1) - 9) p''_0 (x)}{3072 x^4 p_0 (x)^6} 
  + \frac{5 l (l + 1) p_0^{(3)} (x)}{768 x^3 p_0 (x)^6} 
  + \frac{89 l (l + 1) p_0^{(4)} (x)}{12288 x^2 p_0 (x)^6} 
  - \frac{5 p_0^{(6)} (x)}{32768 p_0 (x)^6} \notag \\
  &\quad - \frac{135 l (l + 1) p''_0 (x)^2}{2048 x^2 p_0 (x)^7} 
  - \frac{p_0^{(3)} (x)^2}{2048 p_0 (x)^7} 
  + \frac{105 p_0^{(4)} (x) \, p''_0(x)}{16384 p_0 (x)^7} 
  - \frac{371 p''_0 (x)^3}{16384 p_0 (x)^8} 
  + d (*) , 
\end{align}
Here $d (*)$ denotes a total derivative. 
The odd-order coefficients are total derivatives, while the even-order coefficients give non-trivial contributions to the local WKB periods. 
These coefficients determine the local WKB periods discussed in Section~\ref{subsec:WKB_period}. 

\subsubsection{Diagonalization of the second row}
\label{subsubsec:2nd_diag}
We next consider the diagonalization of the second row. 
Since the diagonal entry $f_2 (x)$ does not depend on $g_{2 ,\, 3} (x)$, its determination at a given order only requires the constraint $r_{21} (x) = 0$. 
The equation $r_{23} (x) = 0$, which fixes $g_{2 ,\, 3} (x)$, can therefore be postponed. 
Thus the computation of the diagonal coefficient $f_2 (x)$ decouples from the determination of $g_{2 ,\, 3}(x)$ at this stage. 
\par
We expand
\begin{equation}
  g_{2 ,\, 1} (x) = \sum_{k = 0}^{\infty} \epsilon^{k + \frac{1}{2}} \, S_{2 ,\, 1}^{(k)} (x) , 
  \qquad
  f_2 (x) = \sum_{k = 0}^{\infty} \epsilon^k \, f_2^{(k)} (x) . 
\end{equation}
Accordingly, we also expand 
\begin{equation}
  r_{21} (x) = \sum_{k = 0}^{\infty} \epsilon^{k + \frac{1}{2}} \, r_{21}^{(k)} (x) . 
\end{equation}
At the lowest orders, the second-row equations take the form 
\begin{align}
  r_{21}^{(0)} &= - \frac{\xi}{x} \, p_0 + p_1 + p_0 S_{2 ,\, 1}^{(0)} 
  - \frac{\xi}{x} \, S_{3 ,\, 1}^{(0)} + S_{2 ,\, 1}^{(0)} S_{3 ,\, 1}^{(0)} = 0 , \\
  f_2^{(0)} &= 2 p_0 , \\
  r_{21}^{(1)} &= \frac{\xi}{x} \, S_{2 ,\, 1}^{(0)} S_{3 ,\, 2}^{(0)} - \frac{l}{x} \, S_{2 ,\, 1}^{(0)} 
  + p_0 S_{2 ,\, 1}^{(1)} + S_{2 ,\, 1}^{(1)} S_{3 ,\, 1}^{(0)} 
  - \frac{\xi}{x} \, S_{3 ,\, 1}^{(1)} + S_{2 ,\, 1}^{(0)} S_{3 ,\, 1}^{(1)} 
  + S_{2 ,\, 1}^{(0) \prime} = 0 , \\
  f_2^{(1)} &= - \frac{\xi}{x} \, S_{2 ,\, 1}^{(0)} 
  - \frac{\xi}{x} \, S_{3 ,\, 2}^{(0)} + S_{2 ,\, 1}^{(0)} S_{3 ,\, 2}^{(0)} .
\end{align}
\par
Using the minus-branch data obtained in Subsection~\ref{subsubsec:3rd_diag}, 
\begin{equation}
  S_{3 ,\, 1}^{(0)} (x) = - p_0 (x), \quad 
  S_{3 ,\, 2}^{(0)} (x) = - \frac{\xi}{x} , 
\end{equation}
together with $p_1 (x) = 0$, we find 
\begin{equation}
  f_2^{(1)} (x) = 0 . 
\end{equation}
Thus, also in the second row, the first non-trivial contribution starts only at order $\epsilon^2$. 
Substituting the third-row coefficients into $r_{21}^{(1)} = 0$, we obtain 
\begin{equation}
  - \frac{S_{2 ,\, 1}^{(0)} (x) \, p'_0 (x)}{2 p_0 (x)}
  + S_{2 ,\, 1}^{(0) \prime} (x)
  - \frac{l \, \xi}{x^2}
  + \frac{\xi \, p'_0 (x)}{2 x \, p_0 (x)} = 0 .
\end{equation}
Multiplying this equation from the left by $\xi$, and using the nilpotency $\xi^2 = 0$, we obtain an ordinary first-order equation for the bosonic combination $\xi \, S_{2 ,\, 1}^{(0)} (x)$: 
\begin{equation}
  - \frac{\xi \, S_{2 ,\, 1}^{(0)} (x) \, p'_0 (x)}{2 p_0 (x)} 
  + \xi \, S_{2 ,\, 1}^{(0)\prime} (x) = 0 . 
\end{equation}
This integrates to 
\begin{equation}
  \xi \, S_{2 ,\, 1}^{(0)} (x) = C \sqrt{p_0 (x)} , 
  \label{eq:S210_sol}
\end{equation}
where $C$ is a Grassmann-even nilpotent integration constant. 
Accordingly, 
\begin{equation}
  f_2^{(2)} (x) = - \frac{C (l + 1)}{2 x^2 \sqrt{p_0 (x)}} . 
  \label{eq:f22}
\end{equation}
\par
This behavior is qualitatively different from that of the third row. 
Unlike the third-row coefficients, which are obtained recursively from local algebraic-differential relations, the second-row coefficient $f_2 (x)$ already requires an integration at the first non-trivial step; see Eq.~\eqref{eq:S210_sol}. 
This essential difference suggests that the second row is governed by a semi-local structure. 
Higher-order coefficients are therefore expected to involve iterated integrals and to be related to semi-local quantities rather than to the local WKB periods. 
\par
This feature is reminiscent of semi-local charges involving inverse derivatives on the cylinder, such as $\tilde{\mathbf{I}}_1$ in Eq.~(4.5) of \cite{alfimov2015}. 
In the NS sector, where the supercurrent has no zero mode, an inverse derivative can be defined mode by mode.
The appearance of the integration constant $C$ in Eq.~\eqref{eq:S210_sol} suggests that the second-row period $Q_2^{(\mathrm{sloc})}$ may be related to such a semi-local structure. 
A direct identification, however, requires a CFT interpretation of the Grassmann-even nilpotent integration constant $C$, which we leave for future work. 

\subsection{WKB periods}
\label{subsec:WKB_period}
For the present $C (2)^{(2)}$ problem, the WKB periods \eqref{eq:WKB_period} associated with the third and second rows are defined by
\begin{equation}
  Q_i^{(\mathrm{loc})} \coloneqq \oint_{\mathcal{C}} f_3^{(i)} (x) \, dx , \quad 
  Q_i^{(\mathrm{sloc})} \coloneqq \oint_{\mathcal{C}} f_2^{(i)} (x) \, dx ,
\end{equation}
where $\mathcal{C}$ is the Pochhammer contour shown in Fig.\,\ref{fig:Pochhammer}. 
Here, $Q_i^{(\mathrm{loc})}$ are the local periods associated with the third row, while $Q_i^{(\mathrm{sloc})}$ denote the semi-local periods associated with the second row. 
\par
Since integrands of the form $p_0 (x)^a \, x^b$ appear repeatedly, it is convenient to introduce the contour integral 
\begin{equation}
  J (a , b) \coloneqq \oint_{\mathcal{C}} p_0 (x)^a \, x^b \, dx 
  = \oint_{\mathcal{C}} \left( x^{\frac{h M}{2}} - 1 \right)^{\! a} \, x^b \, dx 
  = - \frac{e^{\pi i a} \, \pi i}{h M} 
  \frac{\Gamma \left( - a - \frac{2 (b + 1)}{h M} \right)}
  {\Gamma (- a) \, \Gamma \left( 1 - \frac{2 (b + 1)}{h M} \right)} . 
  \label{eq:Jab}
\end{equation}
This definition allows us to express all period integrals in a common basis and to relate different integrals systematically through the recurrence relation below. 
Using $\Gamma (x + 1) = x \, \Gamma (x)$, one finds the recurrence relation 
\begin{equation}
  J \left( a + m ,\, b + \frac{h M n}{2} \right) = e^{\pi i m} \, 
  \frac{\Gamma \left( \frac{2 (b + 1)}{h M} + n \right) \, \Gamma (a + m + 1) \, 
  \Gamma \left( a + \frac{2 (b + 1)}{h M} + 1 \right)}
  {\Gamma \left( a + m + \frac{2 (b + 1)}{h M} + n + 1 \right) \, 
  \Gamma \left( \frac{2 (b + 1)}{h M} \right) \, \Gamma (a + 1)} \, J (a , b) . 
  \label{eq:J_shift}
\end{equation}
For convenience, we write $J (- n , - m) \eqqcolon J_{n ,\, m}$. 
\par
Using the coefficients $f_3^{(k)} (x)$, the local periods up to sixth order are obtained as 
\begin{align}
  Q_2^{(\mathrm{loc})} 
  &= \frac{M J_{2 ,\, 2}}{2 (M + 1)} \left[ l (l + 1) - \frac{M - 1}{8} \right] , \\
  Q_3^{(\mathrm{loc})} &= 0 , \\
  Q_4^{(\mathrm{loc})} 
  &= \frac{M^2 \, J_{5 ,\, 4}}{16 (M + 1) (4 M + 3)} 
  \left[ l (l + 1) (8 l (l + 1) - 5 M - 1) 
  - \frac{(M - 3) (M - 1) (2 M + 1)}{16} \right] 
  , \\
  Q_5^{(\mathrm{loc})} &= 0, \\
  Q_6^{(\mathrm{loc})} 
  &= \frac{7 M^3}{2048 (M + 1) (6 M + 5) (7 M + 5)} \notag \\
  &\quad \times \bigg[ l (l + 1) \big( - 1120 l (l + 1) M + 32 l (l + 1) (24 l (l + 1) - 17) \notag \\
  &\qquad - 46 M^3 + 491 M^2 + 514 M + 121 \big) 
  - \frac{(M - 5) (M - 1) (M (4 M + 5) (6 M - 13) - 27)}{8} \bigg] . 
\end{align}
As expected from the fact that the odd-order coefficients $f_3^{(2 k + 1)} (x)$ are total derivatives, the corresponding odd-order local periods vanish. 
Thus the first non-trivial local periods arise only at even orders. 
Higher-order local periods are summarized in Appendix~\ref{sec:higher_WKB}. 
\par
For later comparison with the SCFT side, it is convenient to introduce the symmetric polynomial 
\begin{equation}
  s_1 \coloneqq \left( \! l + \frac{1}{2} \right)^{\!\! 2} = l (l + 1) + \frac{1}{4} . 
\end{equation}
Then the local periods become polynomials in $s_1$. 
For instance, 
\begin{align}
  Q_2^{(\mathrm{loc})} &= M \, J_{2 ,\, 2} 
  \left( \frac{1}{M + 1} \, s_1 - \frac{1}{8} \right) , 
  \label{eq:Qloc2} \\
  Q_4^{(\mathrm{loc})} &= \frac{M^2 \, J_{5 ,\, 4}}{2 (4 M + 3)} 
  \left( \frac{1}{M + 1} \, s_1^2 - \frac{5}{8} \, s_1 
  - \frac{1}{64} M^2 + \frac{9}{128} M + \frac{9}{128} \right) , 
  \label{eq:Qloc4}\\
  Q_6^{(\mathrm{loc})} &= \frac{21 M^3 \, J_{8 ,\, 6}}{8 (6 M + 5) (7 M + 5)} \notag \\
  &\quad \times \Bigg[
    \frac{1}{M + 1} \, s_1^3 
    - \frac{35}{24} \, s_1^2 
    - \frac{46 M^2 - 537 M - 537}{768} \, s_1 \notag \\
    &\qquad\quad - \frac{M^4}{256} + \frac{95 M^3}{3072} 
    - \frac{95 M^2}{2048} - \frac{475 M}{3072} - \frac{475}{6144} 
  \Bigg] . 
  \label{eq:Qloc6}
\end{align}
\par
On the other hand, the second row gives the low-order semi-local period 
\begin{equation}
  Q_2^{(\mathrm{sloc})} 
  = - \frac{1}{2} C (l + 1) J_{\frac{1}{2} ,\, 2} , 
\end{equation}
where $C$ is a Grassmann-even nilpotent integration constant. 
This already suggests a structure qualitatively different from the local periods arising from the third row. 

%% file: SCFT.tex
\section{Superconformal field theory}
\label{sec:SCFT}
In this section, we use $\left( z , \bar{z} \right)$ for the coordinates on the complex plane and focus on the holomorphic sector\footnote{
  Some of the OPE computations in this section were performed with the Mathematica package OPEdefs \cite{thielemans1991}. 
}. 
The $\mathcal{N} = 1$ superconformal symmetry is generated by the energy-momentum tensor $T (z)$ and the supercurrent $G (z)$, whose OPEs are 
\begin{align}
  T (z) \, T (w) &\sim \frac{\frac{c}{2}}{(z - w)^4} + \frac{2 T (w)}{(z - w)^2} 
  + \frac{\partial_w T (w)}{z - w} + \mathrm{regular} , 
  \label{eq:OPE_TT} \\
  T (z) \, G (w) &\sim + \frac{\frac{3}{2} \, G (w)}{(z - w)^2} 
  + \frac{\partial_w G (w)}{z - w} + \mathrm{regular} , 
  \label{eq:OPE_TG} \\
  G (z) \, G (w) &\sim + \frac{\frac{2 c}{3}}{(z - w)^3} 
  + \frac{2 T (w)}{z - w} + \mathrm{regular} , 
  \label{eq:OPE_GG}
\end{align}
where $c$ is the central charge. 
From these OPEs, one obtains the $\mathcal{N} = 1$ super-Virasoro algebra. 
We expand the holomorphic fields as follows 
\begin{equation}
  T (z) = \sum_{n \in \mathbb{Z}} L_n \, z^{- n - 2} , \quad 
  G (z) = \sum_r G_r \, z^{- r - \frac{3}{2}} , 
\end{equation}
where $r \in \mathbb{Z} + \frac{1}{2}$ in the Neveu--Schwarz (NS) sector and 
$r \in \mathbb{Z}$ in the Ramond (R) sector. 
The modes satisfy super-Virasoro (superconformal) algebra: 
\begin{align}
  [L_m , L_n] &= (m - n) L_{m + n} 
  + \frac{c}{12} \left( m^3 - m \right) \delta_{m + n ,\, 0} , 
  \label{eq:sVir_LL} \\
  [L_m , G_r] &= \left( \frac{m}{2} - r \right) G_{m + r} , 
  \label{eq:sVir_LG} \\
  \{ G_r , G_s \} &= 2 L_{r + s} 
  + \frac{c}{3} \left( \! r^2 - \frac{1}{4} \right) \delta_{r + s ,\, 0} . 
  \label{eq:sVir_GG}
\end{align}
\par
The primary field $\varPhi (z)$ with conformal dimension $\varDelta$ is defined by the OPE 
\begin{equation}
  T (z) \, \varPhi (w) = \frac{\varDelta \varPhi (w)}{(z - w)^2} 
  + \frac{\partial_w \varPhi (w)}{z - w} + \cdots . 
\end{equation}
Let $A (z)$ and $B (z)$ be fields of conformal dimensions $\varDelta_A$ and $\varDelta_B$, respectively. 
We write their operator product expansion in the form
\begin{equation}
  A (z) \, B (w) = \sum_{k = 1}^{\varDelta_A + \varDelta_B} 
  \frac{\{ A B \}_k (w)}{(z - w)^k} + :\! A B \!:\! (w) + \mathcal{O} (z - w) . 
\end{equation}
Here $:\! A B \!:\! (w)$ denotes the normal ordered product on the complex plane, defined in terms of the radial ordered product by
\begin{equation}
  :\! A B \!:\! (w) \coloneqq \frac{1}{2 \pi i} \oint_w dz \, 
  \frac{\mathcal{R} (A (z) \, B (w))}{z - w} . 
  \label{eq:norord_pl}
\end{equation}

\subsection{Free field realization of super-Virasoro algebras}
\label{subsec:free_fld}
In this subsection, we realize the $\mathcal{N} = 1$ super-Virasoro algebra in terms of a free boson and a free fermion. 
We start from the quantum Miura transformation \cite{bershadsky1985}: 
\begin{align}
  T (z) &\coloneqq \frac{1}{2} \!:\! \partial \phi \, \partial \phi \!:\! (z) 
  + \frac{1}{2} \!:\! \psi \, \partial \psi \!:\! (z) 
  - \alpha_0 \rho \, \partial^2 \phi (z) , 
  \label{eq:free_bos} \\
  G (z) &\coloneqq \; :\! \partial \phi \, \psi \!:\! (z) 
  - 2 \alpha_0 \rho \, \partial \psi (z) , 
  \label{eq:free_fer}
\end{align}
where $\phi (z)$ is the free boson and $\psi (z)$ is the Majorana fermion. 
Here, $\alpha_0$ denotes the background-charge parameter. 
For the $\mathcal{N} = 1$ super-Virasoro case, the quantum Miura transformation takes the above form with $\rho = \frac{1}{2}$. 
We define their OPEs by 
\begin{align}
  \phi (z) \, \phi (w) &\sim \log (z - w) , 
  \label{eq:bos_bos} \\
  \phi (z) \, \psi (w) &\sim 0 , 
  \label{eq:bos_fer} \\
  \psi (z) \, \psi (w) &\sim \frac{1}{z - w} . 
  \label{eq:fer_fer}
\end{align}
The central charge is given by 
\begin{equation}
  c = \frac{3}{2} - 12 \alpha_0^2 \rho^2 = \frac{3}{2} - 3 \alpha_0^2 . 
\end{equation}
The free fields are expanded in modes as 
\begin{align}
  \phi (z) &= q + a_0 \log z - \sum_{n \neq 0} \frac{a_n}{n} z^{- n} , 
  \label{eq:bos_mode} \\
  \psi (z) &= \sum_r \psi_r \, z^{- r - \frac{1}{2}} , 
  \label{eq:fer_mode}
\end{align}
where $r \in \mathbb{Z} + \frac{1}{2}$ in the NS sector and $r \in \mathbb{Z}$ in the R sector. 
The modes satisfy the Neveu--Schwarz--Ramond (NSR) algebra: 
\begin{equation}
  [a_m , a_n] = m \, \delta_{m + n ,\, 0} , \quad 
  [a_0 , q] = 1 , \qquad 
  \{ \psi_r , \psi_s \} = \delta_{r + s ,\, 0} . 
\end{equation}
\par
As primary fields corresponding to the NS and R vacua, we introduce the vertex operators 
\begin{equation}
  V_\varLambda (z) = \;:\! e^{\varLambda \phi} \!:\! (z) , \quad 
  W_\varLambda^\pm = \;:\! \sigma^\pm \, e^{\varLambda \phi} \!:\! (z) , 
\end{equation}
where $\sigma^\pm$ are the spin fields associated with the Majorana fermion $\psi$. 
The conformal dimensions of these vertex operators are 
\begin{equation}
  \varDelta_{\mathrm{NS}} = \frac{1}{2} \varLambda (\varLambda + \alpha_0) , \quad 
  \varDelta_{\mathrm{R}} = \varDelta_{\mathrm{NS}} + \frac{1}{16} . 
\end{equation}
Here the shift by $\frac{1}{16}$ in $\varDelta_{\mathrm{R}}$ comes from the conformal weight of the spin field. 
For later convenience, we also introduce the symmetric polynomial 
\begin{equation}
  \sigma_1 \coloneqq \left( \varLambda + \frac{\alpha_0}{2} \right)^{\!\! 2} . 
  \label{eq:sym_pol}
\end{equation}
In terms of $\sigma_1$, the NS conformal dimension is written as 
\begin{equation}
  \varDelta_{\mathrm{NS}} = \frac{1}{2} \sigma_1 - \frac{\alpha_0^2}{8} . 
\end{equation}
\par
In the NS sector, the highest-weight state $\ket{\varLambda}_{\mathrm{NS}}$ is defined by 
\begin{equation}
  a_n \ket{\varLambda}_{\mathrm{NS}} = 0 \quad (n > 0) , \qquad 
  \psi_r \ket{\varLambda}_{\mathrm{NS}} = 0 \quad 
  \left( \! r > 0 ,\; r \in \mathbb{Z} + \frac{1}{2} \right) , 
\end{equation}
together with 
\begin{equation}
  a_0 \ket{\varLambda}_{\mathrm{NS}} = \varLambda \ket{\varLambda}_{\mathrm{NS}} . 
\end{equation}
\par
In the R sector, the fermion is expanded in integer modes, and the presence of the zero mode $\psi_0$ implies that the ground state is two-fold degenerate. 
We denote the two Ramond highest-weight states by $\ket{\varLambda , \pm}_{\mathrm{R}}$, which satisfy 
\begin{equation}
  a_n \ket{\varLambda , \pm}_{\mathrm{R}} 
  = \psi_n \ket{\varLambda , \pm}_{\mathrm{R}} = 0 \quad (n > 0) , 
\end{equation}
and
\begin{equation}
  a_0 \ket{\varLambda , \pm}_{\mathrm{R}} 
  = \varLambda \ket{\varLambda , \pm}_{\mathrm{R}} . 
\end{equation}
The fermionic zero mode acts as
\begin{equation}
  \psi_0 \ket{\varLambda , \pm}_{\mathrm{R}} 
  = \frac{1}{\sqrt{2}} \ket{\varLambda , \mp}_{\mathrm{R}} , 
\end{equation}
which realizes the Clifford algebra in the Ramond ground-state sector. 

\subsection{Local IoMs on the complex plane}
\label{subsec:IoM}
On the complex plane, the local IoMs and the supercharge are defined by contour integrals around the origin 
\begin{equation}
  \bm{H}_{n - 1} \coloneqq \oint \frac{dz}{2 \pi i} \, J_n (z) \quad 
  (n \in \mathbb{Z}) , \qquad 
  \bm{Q} \coloneqq \oint \frac{dz}{2 \pi i} \, G (z) . 
  \label{eq:charge_def}
\end{equation}
Here, $J_n (z)$ denotes the conserved current of spin $n$, which can be constructed from normal ordered products of the energy-momentum tensor $T$, supercurrent $G$ and their derivatives. 
The currents $J_n (z)$ are determined up to total derivative terms by the requirement of mutual commutativity: 
\begin{equation}
  [\bm{H}_m , \bm{H}_n] = 0 \quad 
  (m , n \in \mathbb{Z}) , \qquad 
  [\bm{H}_n , \bm{Q}] = 0 , \quad 
  \{ \bm{Q} , \bm{Q} \} = 2 \bm{H}_1 . 
  \label{eq:conserved_def}
\end{equation}
Since the charges are defined by contour integrals, currents differing by total derivatives give the same conserved charge. 
In the present super-Virasoro case, the odd-spin bosonic currents $J_{2 k + 1} (z)$ reduce to total derivatives and hence do not contribute to \eqref{eq:charge_def}. 
Moreover, imposing the commutativity conditions \eqref{eq:conserved_def}, one finds no additional independent local conserved charges of half-integer spin. 
The local IoMs of the $\mathcal{N} = 1$ super-Virasoro theory have been studied in several equivalent formalisms. 
In particular, the conserved quantities were written in superfield form in \cite{mathieu1990}, while higher-spin local IoMs were constructed explicitly in component form in \cite{yamanaka1988}. 
The first non-trivial currents are given by 
\begin{align}
  J_2 (z) &\coloneqq T (z) , \\
  J_4 (z) &\coloneqq \; :\! T T \!:\! (z) 
  + \frac{1}{4} \!:\! \partial G \, G \!:\! (z) , \\
  J_6 (z) &\coloneqq \; :\!T  \!:\! T T \!::\! (z) 
  - \frac{c + 2}{12} \!:\! \partial T \, \partial T \!:\! (z) 
  + \frac{1}{2} \!:\! T \!:\! \partial G \, G \!::\! (z) 
  - \frac{c - 1}{48} \!:\! \partial^2 G \, \partial G \!:\! (z) . 
\end{align}
Higher-spin currents can be constructed in the same manner. 
In particular, starting from the most general linear combination of normally ordered monomials with the required spin, the coefficients are fixed by imposing the commutativity conditions \eqref{eq:conserved_def}, modulo total derivative terms. 
Explicit component expressions rapidly become lengthy; for this reason we only display the currents needed for the direct zero-mode computation up to sixth order. 
A systematic construction of the higher-spin currents would allow one to extend the direct SCFT-side computation and to compare the resulting eigenvalues with the higher WKB periods listed in Appendix~\ref{sec:higher_WKB}. 

\subsection{Conformal transformation and zero mode formulae on the cylinder}
\label{subsec:conf_transf}
To compare the ODE side with the CFT side, we need to map the conserved currents from the complex plane to the cylinder and evaluate their zero modes. 
In Virasoro CFT, because the normal ordering prescriptions on the plane and on the cylinder are different, non-trivial correction formulae are required; such formulae were developed in \cite{dymarsky2020,novaes2021}. 
However, those results apply only to periodic operators. 
They can therefore be used directly in the R sector, but not in the NS sector, where anti-periodic operators appear. 
In this subsection, we present the corresponding formulae for the anti-periodic case. 
Their derivation will be given in Appendix~\ref{sec:conf_map}. 
\par
We use the conformal map
\begin{equation}
  z = e^u , 
  \label{eq:conf_transf}
\end{equation}
from the cylinder coordinate $u$ to the complex plane coordinate $z$. 
For a primary field $A (z) = \sum \limits_n A_n z^{- n - \varDelta_A}$ of conformal weight $\varDelta_A$, the transformed field on the cylinder is 
\begin{equation}
  \widetilde{A} (u) 
  = \left( \frac{dz}{du} \right)^{\!\! \varDelta_A} A (z) 
  = z^{\varDelta_A} \, A (z) 
  = \sum_{n \in \mathbb{Z}} A_n z^{- n} . 
\end{equation}
For a non-primary field, the transformation rule is written as 
\begin{equation}
  \widetilde{A} (u) 
  = z^{\varDelta_A} \, A (z) + \delta A (z) 
  \eqqcolon \sum_{n \in \mathbb{Z}} \widetilde{A}_n z^{- n} , 
  \label{eq:non-primary}
\end{equation}
where $\delta A (z)$ denotes the extra contribution arising because $A (z)$ is not a primary field. 
In particular, for the energy-momentum tensor one has 
\begin{equation}
  \widetilde{T} (u) = z^2 \, T (z) + \delta T (z) 
  = z^2 \, T (z) - \frac{c}{24} , 
  \label{eq:EM_tensor}
\end{equation}
where $\delta T (z)$ is the contribution from the Schwarzian derivative. 
For later convenience, we introduce 
\begin{equation}
  A_R (z) = A (z) + z^{- \varDelta_A} \, \delta A (z) . 
\end{equation}
\par
Under the conformal transformation \eqref{eq:conf_transf}, a conserved current $J (z)$ is mapped to the corresponding field $\widetilde{J} (u)$ on the cylinder. 
To compute its zero mode, we need the normal ordered products on the cylinder. 
For two fields $\widetilde{A}$ and $\widetilde{B}$, the normal ordered product on the cylinder is defined by the time-ordered product with respect to the cylinder time $\Re u$: 
\begin{equation}
  (\widetilde{A} \widetilde{B}) (v) 
  \coloneqq \oint_v \frac{du}{2 \pi i} 
  \frac{\mathcal{T} \left( \! \widetilde{A} (u) \, \widetilde{B} (v) \! \right)}{u - v} . 
\end{equation}
This differs from the normal ordered product \eqref{eq:norord_pl} on the complex plane, which is defined by radial ordering. 
Using the conformal transformation \eqref{eq:conf_transf} together with \eqref{eq:non-primary}, one finds 
\begin{equation}
  (\widetilde{A} \widetilde{B}) (v) 
  = \frac{1}{2 \pi i} \oint_v \frac{dz}{z} 
  \frac{z^{\varDelta_A} \, w^{\varDelta_B} \, \mathcal{R} (A_R (z) \, B_R (w))}{\log \frac{z}{w}} , 
  \label{eq:norord_cyl}
\end{equation}
which expresses the cylinder normal ordering in terms of operator products on the complex plane. 
This rewriting is particularly useful for higher-spin operators and will allow us to derive explicit formulae for the zero modes on the cylinder. 
\par
For periodic operators, the normal ordered product on the cylinder can be written as 
\begin{align}
  (\widetilde{A} \widetilde{B}) (w) 
  = \widetilde{A}_- (w) \, \widetilde{B} (w) 
  &+ (- 1)^{|A| |B|} \widetilde{B} (w) \, \widetilde{A}_+ (w) \notag \\
  &+ \sum_{k = 1}^{\varDelta_A + \varDelta_B} 
  \left[ b_k (\varDelta_A - 1) - \frac{(\varDelta_A - 1)_k}{k !} \right] 
  \{ A_R B_R \}_k (w) \, w^{\varDelta_A + \varDelta_B - k} , 
\end{align}
where 
\begin{equation}
  \{ A_R B_R \}_k (w) = \sum_m (\{ A_R B_R \}_k)_m \, w^{- \varDelta_A - \varDelta_B + k - m} , 
\end{equation}
$b_k (x)$ denotes the Bernoulli polynomial of the second kind, 
\begin{equation}
  \frac{(t + 1)^x}{\log (t + 1)} = \sum_{n = 0}^{\infty} b_n (x) \, t^{n - 1} , 
  \qquad |t| < 1 , 
  \label{eq:Bernoulli}
\end{equation}
and $(a)_k = a (a - 1) \cdots (a - k + 1)$. 
Accordingly, the zero mode is given by 
\begin{equation}
  (\widetilde{A} \widetilde{B})_0 
  = \sum_{n = 1}^{\infty} \widetilde{A}_{- n} \, \widetilde{B}_n 
  + (- 1)^{|A| |B|} \sum_{n = 0}^{\infty} \widetilde{B}_{- n} \, \widetilde{A}_n 
  + \sum_{k = 1}^{\varDelta_A + \varDelta_B} 
  \left[ b_k (\varDelta_A - 1) - \frac{(\varDelta_A - 1)_k}{k !} \right] 
  (\{ A_R B_R \}_k)_0 . 
\end{equation}
We summarize the derivation of these formulae in Appendix~\ref{subsec:per}. 
\par
In the NS sector, the relevant operators are anti-periodic, and their mode expansions are therefore half-integer moded rather than integer moded. 
Correspondingly, the correction term in the cylinder normal ordering is modified from the periodic case, with $(\varDelta_A - 1)_k$ replaced by $\big( \! \varDelta_A - \frac{1}{2} \big)_{\! k}$. 
As a result, the correction term in the cylinder normal ordering is modified accordingly, and one finds 
\begin{align}
  (\widetilde{A} \widetilde{B}) (v) 
  = \widetilde{A}_- (w) \, \widetilde{B} (w) 
  &+ (- 1)^{|A| |B|} \, \widetilde{B} (w) \, \widetilde{A}_+ (w) \notag \\
  &+ \sum_{k = 1}^{\varDelta_A + \varDelta_B} 
  \left[ b_k (\varDelta_A - 1) 
  - \frac{\big( \! \varDelta_A - \frac{1}{2} \big)_{\! k}}{k !} \right] \{ A_R B_R \}_k (w) \, w^{\varDelta_A + \varDelta_B - k} . 
\end{align}
Accordingly, the zero mode is given by
\begin{equation}
  (\widetilde{A} \widetilde{B})_0 
  = \sum_{r > 0} \widetilde{A}_{- r} \widetilde{B}_r 
  + (- 1)^{|A| |B|} \sum_{r > 0} \widetilde{B}_{- r} \widetilde{A}_r 
  + \sum_{k = 1}^{\varDelta_A + \varDelta_B} 
  \left[ b_k (\varDelta_A - 1) 
  - \frac{\big( \! \varDelta_A - \frac{1}{2} \big)_{\! k}}{k !} \right] 
  (\{ A_R B_R \}_k)_0 , 
  \label{eq:NS_zero_mode}
\end{equation}
with $r \in \mathbb{Z} + \frac{1}{2}$. 
The derivation is summarized in Appendix~\ref{subsec:anti-per}. 

\subsection{Local IoMs and their vacuum eigenvalues on the cylinder}
\label{subsec:IoM_eig}
We now turn to the local conserved currents on the cylinder. 
They are obtained from the corresponding currents on the complex plane by the conformal transformation discussed in the previous subsection. 
For the currents up to spin six, we write 
\begin{align}
  \widetilde{J}_2 (v) &= \widetilde{T} (v) , \\
  \widetilde{J}_4 (v) &= (\widetilde{T} \widetilde{T}) (v) 
  + \frac{1}{4} (\partial \widetilde{G} \, \widetilde{G}) (v) , \\
  \widetilde{J}_6 (v) &= (\widetilde{T} (\widetilde{T} \widetilde{T})) (v) 
  - \frac{c + 2}{12} (\partial \widetilde{T} \, \partial \widetilde{T}) (v) 
  + \frac{1}{2} (\widetilde{T} (\partial \widetilde{G} \, \widetilde{G})) (v) 
  - \frac{c - 1}{48} (\partial^2 \widetilde{G} \, \partial \widetilde{G}) (v) . 
\end{align}
The terms involving only $\widetilde{T}$ and its derivatives were obtained in \cite{bazhanov1996,dymarsky2020,novaes2021}. 
The zero modes of $(\partial \widetilde{G} \, \widetilde{G})$, $(\widetilde{T} (\partial \widetilde{G} \, \widetilde{G}))$, and $(\partial^2 \widetilde{G} \, \partial \widetilde{G})$ are summarized in Appendix~\ref{subsec:zero-mode}. 
Using these ingredients, one can evaluate the zero mode of each cylinder current $\widetilde{J}_{2k}(u)$ by expressing it in terms of cylinder normal ordered products and then applying the formulae of the previous subsection to each term. 
The corresponding conserved charges are obtained by integrating these currents along the spatial circle $u = i \sigma \; (0 \leq \sigma < 2 \pi)$: 
\begin{equation}
  \bm{I}_{2 k - 1} = \int_{0}^{2 \pi} \frac{d \sigma}{2 \pi} \, 
  \widetilde{J}_{2 k} (i \sigma) , \quad 
  \bm{\mathcal{Q}} = \int_{0}^{2 \pi} \frac{d \sigma}{2 \pi} \, 
  \widetilde{G} (i \sigma) . 
\end{equation} 
Substituting the mode expansions 
\begin{equation}
  \widetilde{T} (u) = \sum_n \widetilde{L}_n e^{- n u} , \quad 
  \widetilde{G} (u) = \sum_r \widetilde{G}_r \, e^{- r u} , 
\end{equation}
one obtains each $\bm{I}_{2 k - 1}$ as an operator written in terms of the cylinder modes. 
Our main interest is in the action of these charges on highest-weight states, since their eigenvalues will later be compared with the WKB periods on the ODE side. 
We therefore define 
\begin{equation}
  \bm{I}_{2 k - 1} \ket{\varDelta_{\mathrm{NS}}} 
  = I_{2 k - 1}^{(\mathrm{NS})} \ket{\varDelta_{\mathrm{NS}}} , \quad 
  \bm{I}_{2 k - 1} \ket{\varDelta_{\mathrm{R}}} 
  = I_{2 k - 1}^{(\mathrm{R})} \ket{\varDelta_{\mathrm{R}}} . 
\end{equation}
The explicit expressions for the local IoM eigenvalues in the NS sector up to sixth order are 
\begin{align}
  I_1^{(\mathrm{NS})} &= \varDelta_{\mathrm{NS}} - \frac{c}{24} , \\
  I_3^{(\mathrm{NS})} 
  &= \varDelta_{\mathrm{NS}}^2 
  - \frac{4 c + 9}{48} \, \varDelta_{\mathrm{NS}} 
  + \frac{c (4 c + 21)}{2304} , \\
  I_5^{(\mathrm{NS})}
  &= \varDelta_{\mathrm{NS}}^3 
  - \frac{3 c + 13}{24} \, \varDelta_{\mathrm{NS}}^2 
  + \frac{8 c^2 + 77 c + 135}{1536} \, \varDelta_{\mathrm{NS}} 
  - \frac{c (c + 11) (8 c + 45)}{110592} . 
\end{align}
For comparison with the ODE side, it is convenient to rewrite these eigenvalues in terms of the symmetric polynomial \eqref{eq:sym_pol}. 
In this form, the dependence on the highest-weight is organized in a way that matches naturally with the WKB periods: 
\begin{align}
  I_1^{(\mathrm{NS})} &= \frac{1}{2} \, \sigma_1 - \frac{1}{16} , 
  \label{eq:I2_NS} \\
  I_3^{(\mathrm{NS})} 
  &= \frac{1}{4} \, \sigma_1^2 - \frac{5}{32} \, \sigma_1 
  + \frac{9 - 2 \alpha_0^2}{512} , 
  \label{eq:I4_NS} \\
  I_5^{(\mathrm{NS})}
  &= \frac{1}{8} \, \sigma_1^3 - \frac{35}{192} \, \sigma_1^2 
  - \frac{46 \alpha_0^2 - 537}{6144} \, \sigma_1 
  - \frac{24 \alpha_0^4 - 190 \alpha_0^2 + 475}{49152} . 
  \label{eq:I6_NS}
\end{align}
These expressions agree with the corresponding NS-sector results in \cite{babenko2017}. 
\par
We next turn to the R sector. 
In the R sector, the corresponding eigenvalues are given by 
\begin{align}
  I_1^{(\mathrm{R})} &= \varDelta_{\mathrm{R}} - \frac{c}{24} 
  = \frac{1}{2} \left( \! \varLambda + \frac{\alpha_0}{2} \right)^{\! 2} , 
  \label{eq:I2_R} \\
  I_3^{(\mathrm{R})} 
  &= \varDelta_{\mathrm{R}}^2 
  - \frac{2 c + 3}{24} \, \varDelta_{\mathrm{R}} 
  + \frac{c (c + 3)}{576} , 
  \label{eq:I4_R} \\
  I_5^{(\mathrm{R})} 
  &= \varDelta_{\mathrm{R}}^3 
  - \frac{3 c + 10}{24} \, \varDelta_{\mathrm{R}}^2 
  + \frac{c^2 + 7 c + 33}{192} \, \varDelta_{\mathrm{R}} 
  - \frac{c \left( c^2 + 11 c - 1233 \right)}{13824} . 
  \label{eq:I6_R}
\end{align}
Moreover, in the R sector, the supercurrent possesses a zero mode. 
Accordingly, the supercharge $\bm{\mathcal{Q}}$ acts non-trivially on the Ramond vacuum states: 
\begin{equation}
  \bm{\mathcal{Q}} \ket{\varLambda , \pm} 
  = \frac{1}{\sqrt{2}} \left( \! \varLambda + \frac{\alpha_0}{2} \right) 
  \ket{\varLambda , \mp} . 
\end{equation}
Therefore, on the Ramond vacuum states one finds 
\begin{equation}
  \bm{\mathcal{Q}}^2 \ket{\varLambda , \pm} 
  = \frac{1}{2} \left( \! \varLambda + \frac{\alpha_0}{2} \right)^{\! 2} \ket{\varLambda , \pm} 
  = I_1^{(\mathrm{R})} \ket{\varLambda , \pm} , 
\end{equation}
which is consistent with the super-Virasoro relation $\{ \bm{\mathcal{Q}} , \bm{\mathcal{Q}} \} = 2 \bm{I}_1$. 
This provides a useful consistency check of the cylinder normalization in the R sector. 

%% file: ODEIM.tex
\section{The ODE/IM correspondence for \texorpdfstring{$\bm{C (2)^{(2)}}$}{$C (2)^{(2)}$}-type ODEs}
\label{sec:ODE/IM}
In this section, we study the ODE/IM correspondence for the $C (2)^{(2)}$-type system by comparing the WKB periods obtained in Section~\ref{sec:linear_problem} with the eigenvalues of the local IoMs derived in Section~\ref{sec:SCFT}. 
On the ODE side, we restrict attention to the local periods obtained from the third row. 
On the SCFT side, we compare them with the NS-sector eigenvalues of the local IoMs. 
We therefore restrict ourselves to this sector and summarize the resulting ODE/IM dictionary. 

\subsection*{WKB period \texorpdfstring{$\bm{Q_2}$}{$Q_2$}}
We begin with the lowest non-trivial local WKB period. 
On the ODE side, the second local period is given by 
\begin{equation}
  Q_2^{(\mathrm{loc})} = M \, J_{2 ,\, 2} 
  \left[ \frac{1}{2 (M + 1)} \, s_1 - \frac{1}{16} \right] . 
\end{equation}
On the SCFT side, the corresponding local IoM eigenvalue in the NS sector is 
\begin{equation}
  I_1^{(\mathrm{NS})} = \frac{1}{2} \, \sigma_1 - \frac{1}{16} . 
\end{equation}
The two expressions have exactly the same functional form, provided that we identify 
\begin{equation}
  \sigma_1 = \frac{1}{M + 1} \, s_1 , \quad 
  \varLambda + \frac{\alpha_0}{2} 
  = \frac{1}{\sqrt{M + 1}} \left( \! l + \frac{1}{2} \right) . 
  \label{eq:dictionary1}
\end{equation}
With this identification, the ODE quantity becomes 
\begin{equation}
  Q_2^{(\mathrm{loc})} = M \, J_{2 ,\, 2} \, I_1^{(\mathrm{NS})} . 
\end{equation}
Thus, the second local WKB period is proportional to the corresponding local IoM eigenvalue $I_1^{(\mathrm{NS})}$. 
This provides the basic dictionary that will be used in the higher-order cases below. 

\subsection*{WKB period \texorpdfstring{$\bm{Q_4}$}{$Q_4$}}
Since the third-order local WKB period vanishes, we next consider the fourth-order one. 
At this order, the comparison not only tests the identification \eqref{eq:dictionary1}, but also fixes the relation between the SCFT parameter $\alpha_0$ and the ODE parameter $M$. 
Using Eq.~\eqref{eq:dictionary1}, the WKB period $Q_4$ in Eq.~\eqref{eq:Qloc4} is rewritten as 
\begin{equation}
  Q_4^{(\mathrm{loc})} = \frac{2 (M + 1) M^2 \, J_{5 ,\, 4}}{4 M + 3} 
  \left[ \frac{1}{4} \, \sigma_1^2 - \frac{5}{32} \, \sigma_1 
  + \frac{- 2 M^2 + 9 M + 9}{512 (M + 1)} \right] . 
\end{equation}
On the SCFT side, the corresponding local IoM eigenvalue in the NS sector is 
\begin{equation}
  I_3^{(\mathrm{NS})} = \frac{1}{4} \, \sigma_1^2 - \frac{5}{32} \, \sigma_1 + \frac{9 - 2 \alpha_0^2}{512} . 
\end{equation}
The quadratic and linear terms in $\sigma_1$ already agree as a consequence of \eqref{eq:dictionary1}. 
Requiring the constant terms to coincide further determines the parameter relation 
\begin{equation}
  \alpha_0 = \frac{M}{\sqrt{M + 1}} .
  \label{eq:dictionary2}
\end{equation}
This is the same relation as the one needed for the higher-order $Q_k$ and $I_{k - 1}$ to agree in the $A_r^{(1)}$, $D_r^{(1)}$ and $E_6^{(1)}$ cases \cite{ito2025a,ide2026}. 
With this identification, the fourth local WKB period becomes 
\begin{equation}
  Q_4^{(\mathrm{loc})} = \frac{2 (M + 1) M^2 \, J_{5 ,\, 4}}{4 M + 3} \, I_3^{(\mathrm{NS})} .
\end{equation}
Thus, the fourth local WKB period is proportional to the corresponding local IoM eigenvalue $I_3^{(\mathrm{NS})}$. 
The comparison at this order fixes the remaining parameter dictionary that will be used in the higher-order cases below. 

\subsection*{WKB period \texorpdfstring{$\bm{Q_6}$}{$Q_6$}}
Since the fifth-order local WKB period vanishes, we next consider the sixth-order one. 
At this stage, the identifications \eqref{eq:dictionary1} and \eqref{eq:dictionary2} are already fixed from the comparison of the second- and fourth-order quantities. 
The sixth-order comparison therefore provides a non-trivial test of whether the same dictionary continues to hold for higher local charges. 
Using the identifications \eqref{eq:dictionary1} and \eqref{eq:dictionary2}, the WKB period $Q_6$ in Eq.~\eqref{eq:Qloc6} is rewritten as 
\begin{equation}
  Q^{(\mathrm{loc})}_6 
  = \frac{21 (M + 1)^2 \, M^3 \, J_{8 ,\, 6}}{(6 M + 5) (7 M + 5)} \, I_5^{(\mathrm{NS})} . 
\end{equation}
Thus, the sixth local WKB period is proportional to the corresponding local IoM eigenvalue $I_5^{(\mathrm{NS})}$. 
In particular, the agreement at this order is not used to determine any new parameter relation, but rather serves as a genuine higher-order check of the correspondence. 
Together, these comparisons consistently reproduce the first three non-trivial local IoM eigenvalues in the NS sector. 
This provides strong evidence for the ODE/IM correspondence between the third-row local WKB periods of the $C (2)^{(2)}$-type ODE and the NS-sector local IoMs of the $\mathcal{N} = 1$ super-Virasoro CFT. 

%% file: Conclusion.tex
\section{Conclusions and Discussion}
\label{sec:conclusion}
In this paper, we studied the ODE/IM correspondence for the $C (2)^{(2)}$-type system and its relation to the $\mathcal{N} = 1$ superconformal field theory. 
Our main purpose was to formulate the supersymmetric ODE side without taking the bosonic limit, and to compare the resulting WKB periods with the eigenvalues of local integrals of motion on the SCFT side. 
\par
On the ODE side, we generalized the modified supersymmetric affine Toda field equations and the associated linear problem of \cite{ito2022a}, and proposed boundary conditions better adapted to the conformal limit. 
By diagonalizing the full $C (2)^{(2)}$-type linear problem obtained from this setup, we recovered the local WKB periods associated with the third row, corresponding to the WKB quantities studied in \cite{babenko2017,babenko2019a}, and extended the computation from eighth to tenth order. 
In addition, the analysis of the second row led to an extra structure that was not identified in \cite{ito2022a}, and low-order computations suggest that it is related to semi-local integrals of motion. 
In this sense, our ODE analysis is based on the direct diagonalization of the full $C (2)^{(2)}$-type linear problem rather than on a reduced second-order ODE alone.
\par
On the SCFT side, we derived the conformal-transformation formula from the plane to the cylinder for the anti-periodic case, extending the periodic formula of \cite{novaes2021}. 
Using this result, we computed the eigenvalues of the local integrals of motion on the cylinder in both the NS and R sectors. 
While the integrable structure of the $\mathcal{N} = 1$ theory was formulated in \cite{kulish2004,kulish2005}, and the eigenvalues of the IoMs were previously investigated numerically through the Suzuki equation in \cite{babenko2017}, we derived explicit analytic expressions for the cylinder eigenvalues in the NS and R sectors and used them for a direct comparison with the ODE side. 
We then confirmed that the NS-sector eigenvalues of the cylinder IoMs agree with the local WKB periods of the $C (2)^{(2)}$-type linear problem. 
In this way, we obtained an explicit ODE/IM dictionary for the NS sector based on the diagonalization of the full linear problem. 
\par
There remain several interesting directions for future work. 
First, although the second-row analysis suggests the presence of semi-local conserved quantities, their precise counterpart on the IM side has not yet been identified. 
Second, while the R-sector eigenvalues of the cylinder IoMs were obtained explicitly, the corresponding quantities on the ODE side are still unknown. 
A plausible direction is to modify the boundary conditions of the linear problem by introducing an appropriate twist, so that the monodromy of the wave function reflects the Ramond boundary condition. 
Whether such a modification reproduces the R-sector spectrum remains an interesting problem for future investigation. 
Finally, the present method is expected to apply more broadly to quantum integrable models associated with other affine Lie superalgebras, such as those studied in \cite{tsuboi1997,tsuboi1999a,tsuboi1999,tsuboi2024}. 

%% file: WKB_period.tex
\section{Higher-order WKB periods}
\label{sec:higher_WKB}
In this appendix, we summarize the higher-order local WKB periods associated with the $C (2)^{(2)}$-type ODE. 
For convenience, we present the results directly in terms of the symmetric variable
\begin{equation}
  s_1 \coloneqq \left( \! l + \frac{1}{2} \right)^{\!\! 2} . 
\end{equation}
Since the odd-order local periods vanish, we record here the next two non-trivial even-order cases, $Q_8^{(\mathrm{loc})}$ and $Q_{10}^{(\mathrm{loc})}$. 
The former is consistent with the quantity denoted by $i_7$ in \cite[Eq.~(4.125)]{babenko2019a}, and the lower-order cases are likewise compatible with \cite[Eq.~(4.6)]{babenko2017}. 
To the best of our knowledge, the expression for $Q_{10}^{(\mathrm{loc})}$ is new. 
\begin{align}
  Q_8^{(\mathrm{loc})} 
  &= \frac{225 M^4 \, J_{11 ,\, 8}}{8 (8 M + 7) (9 M + 7) (10 M + 7)} \notag \\
  &\quad \times \Bigg[
    \frac{1}{M + 1} \, s_1^4 
    - \frac{21}{8} \, s_1^3 
    - \frac{7 \left(202 M^2 - 3881 M - 3881 \right)}{9600} \, s_1^2 \notag \\
  &\qquad - \frac{\left( 3800 M^4 - 57734 M^3 + 244143 M^2 + 603754 M + 301877 \right)}{230400} \, s_1 \notag \\
  &\qquad - \frac{17 M^6}{10240} + \frac{7483 M^5}{368640} 
  - \frac{77 M^4}{1200} - \frac{155687 M^3}{7372800}
  + \frac{2646679 M^2}{7372800} + \frac{1089809 M}{2457600} 
  + \frac{1089809}{7372800} 
  \Bigg] , 
  \label{eq:Qloc8}
\end{align}
\begin{align}
  Q_{10}^{(\mathrm{loc})} 
  &= \frac{5005 M^5 \, J_{14 ,\, 10}}{32 (10 M + 9) (11 M + 9) (12 M + 9) (13 M + 9)} \notag \\
  &\quad \times \Bigg[
    \frac{1}{M + 1} \, s_1^5 - \frac{33}{8} \, s_1^4 
    - \frac{94 M^2 - 2517 M - 2517}{320} \, s_1^3 \notag \\
  &\qquad - \frac{\left(5720 M^4 - 129566 M^3 + 883337 M^2 + 2025806 M + 1012903 \right)}{125440} \, s_1^2 \notag \\
  &\qquad - \frac{3}{16056320} \big( 42896 M^6 - 851444 M^5 + 5499576 M^4 -7884971 M^3  \notag \\
  &\qquad\qquad\qquad - 55410013 M^2 - 61761033 M - 20587011 \big) s_1 \notag \\
  &\qquad - \frac{31 M^8}{28672} + \frac{3057 M^7}{163840} 
  - \frac{3072591 M^6}{32112640} + \frac{3272469 M^5}{32112640} 
  + \frac{11075703 M^4}{25690112} \notag \\
  &\qquad\quad - \frac{6369597 M^3}{8028160} - \frac{150747129 M^2}{64225280} 
  - \frac{57326373 M}{32112640} - \frac{57326373}{128450560} \Bigg] . 
  \label{eq:Qloc10}
\end{align}

%% file: conf_map.tex
\section{Normal ordering on the cylinder}
\label{sec:conf_map}
In this appendix, we derive the zero mode formulas for normal ordered products on the cylinder starting from Eq.~\eqref{eq:norord_cyl}. 
\par
Let 
\begin{equation}
  t \coloneqq \frac{z - w}{w} , \quad z = w (1 + t) . 
\end{equation}
Then
\begin{equation}
  \frac{1}{z} \frac{z^{\varDelta_A} \, w^{\varDelta_B}}{\log \frac{z}{w}}
  = w^{\varDelta_A + \varDelta_B - 1} \, 
  \frac{(1 + t)^{\varDelta_A - 1}}{\log (1 + t)} . 
\end{equation}
Using the Bernoulli polynomials of the second kind $b_n (x)$ defined by Eq.~\eqref{eq:Bernoulli}, 
\begin{equation}
  \frac{(1 + t)^x}{\log (1 + t)} 
  = \sum_{n = 0}^{\infty} b_n (x) \, t^{n - 1} , 
  \label{eq:Bernoulli2}
\end{equation}
together with the OPE 
\begin{equation}
  A_R (z) \, B_R (w) 
  = \sum_{k = 1}^{\varDelta_A + \varDelta_B} 
  \frac{\{ A_R B_R \}_k (w)}{(z - w)^k} + \{ A_R B_R \}_0 (w) + \cdots , 
  \label{eq:OPE}
\end{equation}
we expand the integrand of Eq.~\eqref{eq:norord_cyl} around $z = w$. 
The derivations below are obtained by substituting these expansions into Eq.~\eqref{eq:norord_cyl} and taking the residue at $z = w$, with the singular part specified according to the boundary condition. 

\subsection{Periodic operators}
\label{subsec:per}
We first consider the case of periodic operators, following the method of \cite{novaes2021}. 
In this case, the cylinder normal ordered product is governed by the same singular kernel $\frac{1}{z - w}$ as in the periodic sector. 
Substituting Eqs.~\eqref{eq:Bernoulli2} and \eqref{eq:OPE} into Eq.~\eqref{eq:norord_cyl} and taking the residue at $z = w$, only the terms for which the power of $(z - w)$ is $- 1$ survive, namely those with $n = k$. 
The singular part reproduces the mode decomposition 
\begin{equation}
  \widetilde{A}_- (w) \, \widetilde{B} (w) 
  + (- 1)^{|A| |B|} \, \widetilde{B} (w) \, \widetilde{A}_+ (w) , 
\end{equation}
while the regular part gives the correction terms proportional to $\{ A_R B_R \}_k$. 
The subtraction term in Eq.~\eqref{eq:norord_cyl} contributes the coefficient 
$\frac{(\varDelta_A - 1)_k}{k !}$. 
Hence, 
\begin{align}
  (\widetilde{A} \widetilde{B}) (v) 
  = \widetilde{A}_- (w) \, \widetilde{B} (w) 
  &+ (- 1)^{|A| |B|} \widetilde{B} (w) \, \widetilde{A}_+ (w) \notag \\
  &+ \sum_{k = 1}^{\varDelta_A + \varDelta_B} 
  \left[ b_k (\varDelta_A - 1) - \frac{(\varDelta_A - 1)_k}{k !} \right] 
  \{ A_R B_R \}_k (w) \, w^{\varDelta_A + \varDelta_B - k} . 
  \label{eq:nor_per}
\end{align}
Accordingly, the zero mode is given by
\begin{align}
  (\widetilde{A} \widetilde{B})_0 
  = \sum_{n = 1}^{\infty} \widetilde{A}_{- n} \, \widetilde{B}_n 
  &+ (- 1)^{|A| |B|} \sum_{n = 0}^{\infty} 
  \widetilde{B}_{- n} \, \widetilde{A}_n \notag \\
  &+ \sum_{k = 1}^{\varDelta_A + \varDelta_B} \left[ b_k (\varDelta_A - 1) - \frac{(\varDelta_A - 1)_k}{k !} \right] (\{ A_R B_R \}_k)_0 . 
  \label{eq:zero_per}
\end{align}
\par
For later applications, it is convenient to rewrite the first two terms in Eq.~\eqref{eq:nor_per} directly in terms of the normal ordered product on the complex plane. 
For periodic operators, the mode expansions are integer-moded, and the contribution from the lower and upper modes can be expressed in a closed local form. 
More precisely, one obtains 
\begin{align}
  &\quad\; \widetilde{A}_- (w) \, \widetilde{B} (w) 
  + (- 1)^{|A| |B|} \widetilde{B} (w) \, \widetilde{A}_+ (w) \notag \\
  &= w^{\varDelta_A + \varDelta_B} \, \{ A_R B_R \}_0 (w) 
  + \sum_{k = 1}^{\varDelta_A + \varDelta_B} w^{\varDelta_A + \varDelta_B - k} \, 
  \frac{(\varDelta_A - 1)_k}{k !} \, \{ A_R B_R \}_k (w) . 
  \label{eq:ABBA_per}
\end{align}
This is obtained by substituting the OPE of $A_R (z) \, B_R (w)$ into the residue expression for $\widetilde{A}_- (w) \, \widetilde{B} (w) + (- 1)^{|A| |B|} \, \widetilde{B} (w) \, \widetilde{A}_+ (w)$ and expanding $z^{\varDelta_A - 1}$ around $z = w$. 
Substituting Eq.~\eqref{eq:ABBA_per} into Eq.~\eqref{eq:nor_per}, we obtain the simplified expression 
\begin{equation}
  (\widetilde{A} \widetilde{B}) (v) 
  = w^{\varDelta_A + \varDelta_B} \, \{ A_R B_R \}_0 (w) 
  + \sum_{k = 1}^{\varDelta_A + \varDelta_B} 
  w^{\varDelta_A + \varDelta_B - k} \, 
  b_k (\varDelta_A - 1) \, 
  \{ A_R B_R \}_k (w) . 
  \label{eq:nor_per_simplified}
\end{equation}

\subsection{Anti-periodic operators}
\label{subsec:anti-per}
We next consider the case of anti-periodic operators. 
In this case, the singular part of the kernel is modified from 
\begin{equation}
  \frac{1}{z - w} \qquad \text{to} \qquad \frac{\sqrt{z w}}{z - w} . 
\end{equation}
Equivalently, in terms of the variable $t = \frac{z - w}{w}$, the singular part is written as 
\begin{equation}
  \frac{1}{w} \frac{(1 + t)^{\varDelta_A - \frac{1}{2}}}{t} . 
\end{equation}
Therefore, compared with the periodic case, the only modification is the subtraction of this singular kernel. 
Using Eq.~\eqref{eq:Bernoulli2}, we obtain 
\begin{equation}
  \frac{1}{z} \frac{(1 + t)^{\varDelta_A}}{\log (1 + t)} 
  = \frac{1}{w} \frac{(1 + t)^{\varDelta_A - \frac{1}{2}}}{t} 
  + \frac{1}{w} \sum_{n = 1}^{\infty} 
  \left[ b_n (\varDelta_A - 1) 
  - \frac{\big( \varDelta_A - \frac{1}{2} \big)_{\! n}}{n !} \right] t^{n - 1} . 
  \label{eq:NS_kernel_expansion}
\end{equation}
Substituting this expansion together with Eq.~\eqref{eq:OPE} into Eq.~\eqref{eq:norord_cyl} and taking the residue at $z = w$, we again find that only the terms with $n = k$ contribute. 
The singular part now reproduces the half-integer mode decomposition 
\begin{equation}
  \widetilde{A}_- (w) \, \widetilde{B} (w) 
  + (- 1)^{|A| |B|} \, \widetilde{B} (w) \, \widetilde{A}_+ (w) , 
\end{equation}
while the regular part gives the correction terms proportional to $\{A_R B_R\}_k$. 
Hence,
\begin{align}
  (\widetilde{A} \widetilde{B}) (v) 
  = \widetilde{A}_- (w) \, \widetilde{B} (w) 
  &+ (- 1)^{|A| |B|} \, \widetilde{B} (w) \, \widetilde{A}_+ (w) \notag \\
  &+ \sum_{k = 1}^{\varDelta_A + \varDelta_B} 
  \left[ b_k (\varDelta_A - 1) 
  - \frac{\big( \varDelta_A - \frac{1}{2} \big)_{\! k}}{k !} \right] 
  \{ A_R B_R \}_k (w) \, w^{\varDelta_A + \varDelta_B - k} . 
  \label{eq:nor_aper}
\end{align}
Accordingly, the zero mode is given by
\begin{align}
  (\widetilde{A} \widetilde{B})_0 
  = \sum_{r > 0} \widetilde{A}_{- r} \widetilde{B}_r 
  &+ (- 1)^{|A| |B|} \sum_{r \geq 0} \widetilde{B}_{- r} \widetilde{A}_r \notag \\
  &+ \sum_{k = 1}^{\varDelta_A + \varDelta_B} 
  \left[ b_k (\varDelta_A - 1) 
  - \frac{\big( \varDelta_A - \frac{1}{2} \big)_{\! k}}{k !} \right] 
  (\{ A_R B_R \}_k)_0 , 
  \label{eq:zero_aper}
\end{align}
with $r \in \mathbb{Z} + \frac{1}{2}$. 
\par
In the anti-periodic case, one can similarly rewrite the first two terms in Eq.~\eqref{eq:nor_aper} directly in terms of the normal ordered product on the complex plane. 
Because the mode expansions are now half-integer-moded, the resulting coefficient is shifted accordingly. 
More precisely, one finds 
\begin{align}
  &\quad\; \widetilde{A}_- (w) \, \widetilde{B} (w) 
  + (- 1)^{|A| |B|} \, \widetilde{B} (w) \, \widetilde{A}_+ (w) \notag \\
  &= w^{\varDelta_A + \varDelta_B} \, \{ A_R B_R \}_0 (w) 
  + \sum_{k = 1}^{\varDelta_A + \varDelta_B} 
  w^{\varDelta_A + \varDelta_B - k} \, 
  \frac{\bigl( \varDelta_A - \frac{1}{2} \bigr)_{\! k}}{k !} \, 
  \{ A_R B_R \}_k (w) . 
  \label{eq:ABBA_aper}
\end{align}
This is obtained by substituting the OPE of $A_R (z) \, B_R (w)$ into the residue expression for $\widetilde{A}_- (w) \, \widetilde{B} (w) 
+ (- 1)^{|A| |B|} \, \widetilde{B} (w) \, \widetilde{A}_+ (w)$ and expanding $z^{\varDelta_A - \frac{1}{2}}$ around $z = w$. 
Compared with the periodic case, the half-integer moding shifts the coefficient from $\dfrac{(\varDelta_A - 1)_k}{k !}$ to $\dfrac{\bigl( \varDelta_A - \frac{1}{2} \bigr)_{\! k}}{k !}$. 
Substituting Eq.~\eqref{eq:ABBA_aper} into Eq.~\eqref{eq:nor_aper}, we obtain the simplified expression 
\begin{equation}
  (\widetilde{A} \widetilde{B}) (v) 
  = w^{\varDelta_A + \varDelta_B} \, \{ A_R B_R \}_0 (w) 
  + \sum_{k = 1}^{\varDelta_A + \varDelta_B} 
  w^{\varDelta_A + \varDelta_B - k} \, 
  b_k (\varDelta_A - 1) \, 
  \{ A_R B_R \}_k (w) . 
  \label{eq:nor_aper_simplified}
\end{equation}
In particular, the cylinder normal ordered product itself is sector-independent; 
the difference between the periodic and anti-periodic cases appears only in the intermediate mode decomposition and in the corresponding zero mode formulae. 

\subsection{Examples of zero mode computations}
\label{subsec:zero-mode}
In this subsection, we present several explicit examples of zero mode computations for normal ordered products on the cylinder. 
Our main purpose is to show how the general formulae in Appendices \ref{subsec:per} and \ref{subsec:anti-per} are applied in concrete cases, especially for composite operators containing the supercurrent $G$. 
As the simplest example, we begin with the fermionic part of the energy-momentum tensor $T_{\mathrm{fer}}$ and then consider several normal ordered products involving $G$. 
These results will be used in the evaluation of the zero modes of the local IoMs. 
\par
Before turning to explicit examples, let us introduce a compact notation for the correction terms appearing in the periodic and anti-periodic cases. 
We define
\begin{equation}
  f_k^{(\mathrm{per})} (x) \coloneqq b_k (x) - \frac{(x)_k}{k !} , \quad 
  f_k^{(\mathrm{aper})} (x) \coloneqq b_k (x) 
  - \frac{\bigl( x + \frac{1}{2} \bigr)_{\! k}}{k !} . 
\end{equation}
For convenience, we collect below the values that will be used repeatedly in the following examples. 
\begin{table}[H]
  \centering
  \renewcommand{\arraystretch}{1.2}
  \begin{tabular}{c|cccccc}
    $k$ & $1$ & $2$ & $3$ & $4$ & $5$ & $6$ \\
    \hline
    $b_k \big( \frac{1}{2} \big)$ 
    & $1$ & $\frac{1}{24}$ & $0$ & $-\frac{17}{5760}$ & $\frac{17}{5760}$ & $-\frac{2489}{967680}$ \\
    $b_k \big( \frac{3}{2} \big)$ 
    & $2$ & $\frac{25}{24}$ & $\frac{1}{24}$ & $-\frac{17}{5760}$ & $0$ & $\frac{367}{967680}$ 
  \end{tabular}
  \caption{Values of the Bernoulli polynomial of the second kind $b_k (x)$ used in the examples. }
  \label{tab:Bernoulli-values}
\end{table}
\begin{table}[H]
  \centering
  \renewcommand{\arraystretch}{1.2}
  \begin{tabular}{c|cccccc}
    $k$ & $1$ & $2$ & $3$ & $4$ & $5$ & $6$ \\
    \hline
    \quad & \quad & \quad & \quad & \quad & \quad & \quad \\[-6mm]
    $f_k^{(\mathrm{per})} (1)$
    & $\frac{1}{2}$ & $\frac{5}{12}$ & $-\frac{1}{24}$ & $\frac{11}{720}$ & $-\frac{11}{1440}$ & $\frac{271}{60480}$ \\[0.5mm]
    \hline
    \quad & \quad & \quad & \quad & \quad & \quad & \quad \\[-6mm]
    $f_k^{(\mathrm{per})} \big( \frac{1}{2} \big)$
    & $\frac{1}{2}$ & $\frac{1}{6}$ & $-\frac{1}{16}$ & $\frac{13}{360}$ & $-\frac{281}{11520}$ & $\frac{4339}{241920}$ \\
    $f_k^{(\mathrm{per})} \big( \frac{3}{2} \big)$
    & $\frac{1}{2}$ & $\frac{2}{3}$ & $\frac{5}{48}$ & $-\frac{19}{720}$ & $\frac{3}{256}$ & $-\frac{781}{120960}$ \\
    $f_k^{(\mathrm{per})} \big( \frac{5}{2} \big)$
    & $\frac{1}{2}$ & $\frac{7}{6}$ & $\frac{37}{48}$ & $\frac{7}{90}$ & $-\frac{169}{11520}$ & $\frac{1273}{241920}$ 
  \end{tabular}
  \caption{Values of the periodic correction terms 
  $f_k^{(\mathrm{per})} (x) = b_k (x) - \frac{(x)_k}{k !}$ used in the examples.}
  \label{tab:per-corrections}
\end{table}
\begin{table}[H]
  \centering
  \renewcommand{\arraystretch}{1.2}
  \begin{tabular}{c|cccccc}
    $k$ & $1$ & $2$ & $3$ & $4$ & $5$ & $6$ \\
    \hline
    \quad & \quad & \quad & \quad & \quad & \quad & \quad \\[-6mm]
    $f_k^{(\mathrm{aper})} \big( \frac{1}{2} \big)$
    & $0$ & $\frac{1}{24}$ & $0$ & $-\frac{17}{5760}$ & $\frac{17}{5760}$ & $-\frac{2489}{967680}$ \\
    $f_k^{(\mathrm{aper})} \big( \frac{3}{2} \big)$
    & $0$ & $\frac{1}{24}$ & $\frac{1}{24}$ & $-\frac{17}{5760}$ & $0$ & $\frac{367}{967680}$ \\
    $f_k^{(\mathrm{aper})} \big( \frac{5}{2} \big)$
    & $0$ & $\frac{1}{24}$ & $\frac{1}{12}$ & $\frac{223}{5760}$ & $-\frac{17}{5760}$ & $\frac{367}{967680}$ 
  \end{tabular}
  \caption{Values of the anti-periodic correction terms 
  $f_k^{(\mathrm{aper})} (x) = b_k (x) - \dfrac{(x + \frac{1}{2})_k}{k !}$ used in the examples.}
  \label{tab:aper-corrections}
\end{table}

\subsubsection*{Energy-momentum tensor \texorpdfstring{$\bm{\widetilde{T}_{\mathrm{fer}}}$}{$T_{\mathrm{fer}}$}}
We begin with the fermionic part of the energy-momentum tensor: 
\begin{equation}
  T_{\mathrm{fer}} (z) = \frac{1}{2} \!:\! \partial \psi \, \psi \!:\! (z) . 
  \label{eq:T_fer}
\end{equation}
This example illustrates in the simplest setting how the general formulae for cylinder normal ordered products and their zero modes are applied in practice. 
In particular, it makes clear how sector-dependent constant terms arise when one takes the zero mode on the cylinder. 
\par
To compute the cylinder expression corresponding to $T_{\mathrm{fer}}$, we apply the simplified normal-ordering formulae \eqref{eq:nor_aper_simplified} and \eqref{eq:nor_per_simplified} to the pair 
\begin{equation}
  A (w) = \partial_w \psi (w) , \quad B (w) = \psi (w) . 
\end{equation}
Since both fields are fermionic, the resulting normal ordered product on the cylinder is written with parentheses, in contrast to the plane normal ordering denoted by colons. 
In our conventions, the corresponding cylinder fields are 
\begin{equation}
  \widetilde{\psi} (w) = w^{\frac{1}{2}} \, \psi_R (w) , \quad
  \partial \widetilde{\psi} (w) 
  = w^{- \frac{3}{2}} \, \partial_w \left[ w^{\frac{1}{2}} \, \psi_R (w) \right] . 
\end{equation}
For the plane OPE, the only singular coefficients needed here are
\begin{equation}
  \{ \partial \psi_R \psi_R \}_1 (w) = \frac{1}{2 w} , \quad 
  \{ \partial \psi_R \psi_R \}_2 (w) = - 1 . 
\end{equation}
Substituting these into \eqref{eq:nor_per_simplified} or equivalently \eqref{eq:nor_aper_simplified}, one finds 
\begin{equation}
  (\partial \widetilde{\psi} \, \widetilde{\psi}) (v) 
  = w^2 \, \{ \partial \psi \, \psi \}_0 (w) + \frac{11}{24} . 
\end{equation}
Thus, when written in this local form, the cylinder normal ordered product itself is independent of the sector. 
\par
We next take the zero mode by using \eqref{eq:zero_per} and \eqref{eq:zero_aper}. 
At this stage, the periodic and anti-periodic cases differ through the correction terms
$f_k^{(\mathrm{per})}$ and $f_k^{(\mathrm{aper})}$; 
using the values listed in Tables~\ref{tab:per-corrections} and \ref{tab:aper-corrections}, we obtain 
\begin{equation}
  (\partial \widetilde{\psi} \, \widetilde{\psi})_0 = 
  \begin{cases}
    2 \sum \limits_{\substack{r > 0 \\ r \in \mathbb{Z} + \frac{1}{2}}} 
    r \, \widetilde{\psi}_{- r} \widetilde{\psi}_r - \dfrac{1}{24} 
    & (\mathrm{NS}) , \\[1.2ex]
    2 \sum \limits_{\substack{r > 0 \\ r \in \mathbb{Z}}} 
    r \, \widetilde{\psi}_{- r} \widetilde{\psi}_r + \dfrac{1}{12} 
    & (\mathrm{R}) . 
  \end{cases}
\end{equation}
Here we used the Ramond zero mode relation $\{ \psi_0 , \psi_0 \} = 1$, which implies $\psi_0^2 = \frac{1}{2}$. 
Using Eq.~\eqref{eq:T_fer}, this immediately gives the corresponding zero mode contribution of the fermionic energy-momentum tensor. 
\par
The constant term in the NS sector is precisely the Schwarzian contribution $- \frac{c}{24}$ associated with the conformal transformation from the plane to the cylinder, with the central charge of the free fermion 
\begin{equation}
  c_{\mathrm{fer}} = \frac{1}{2} . 
\end{equation}
Moreover, the difference between the Ramond and Neveu--Schwarz constants is 
\begin{equation}
  \frac{1}{2} \left[ \frac{1}{12} - \left( \! - \frac{1}{24} \right) \! \right] 
  = \frac{1}{16} , 
\end{equation}
which reproduces the familiar conformal weight shift between the Ramond ground state and the Neveu--Schwarz vacuum. 
In this way, the general formulae correctly recover both the Schwarzian contribution and the expected Ramond--Neveu--Schwarz offset already in the simplest example. 

\subsubsection*{Spin-4: \texorpdfstring{$\bm{(\partial \widetilde{G} \, \widetilde{G})}$}{$(\partial G \, G)$}}
As the next example, we consider the spin-4 composite field built from the supercurrent. 
Applying Eq.~\eqref{eq:nor_per_simplified} to the case 
\begin{equation}
  A (w) = \partial_w G (w) , \quad 
  B (w) = G (w) , 
\end{equation}
we obtain the cylinder expression for the corresponding normal ordered product. 
In this case, the OPE coefficients entering the correction terms are 
\begin{equation}
  \begin{aligned}
    \{ \partial G_R \, G_R \}_1 (w) &= \frac{c}{w^3} + \frac{3}{w} \, T (w) , & 
    \{ \partial G_R \, G_R \}_2 (w) &= - 2 T (w) - \frac{c}{w^2} , \\
    \{ \partial G_R \, G_R \}_3 (w) &= \frac{c}{w} , & 
    \{ \partial G_R \, G_R \}_4 (w) &= - 2 c . 
  \end{aligned}
\end{equation}
Substituting these into Eq.~\eqref{eq:nor_per_simplified}, one finds 
\begin{equation}
  (\partial \widetilde{G} \, \widetilde{G}) (v) 
  = w^4 \, \{ \partial G \, G \}_0 (w) + \frac{3}{2} \, w^3 \, \partial T 
  + \frac{11}{12} \, w^2 \, T + \frac{17 c}{2880} 
\end{equation}
Next, using Eqs.\,\eqref{eq:zero_per} and \eqref{eq:zero_aper}, together with the coefficients listed in Tables~\ref{tab:per-corrections} and \ref{tab:aper-corrections}, we obtain the zero modes 
\begin{equation}
  (\partial \widetilde{G} \, \widetilde{G})_0 = 
  \begin{cases}
    2 \sum \limits_{\substack{r > 0 \\ r \in \mathbb{Z} + \frac{1}{2}}} 
    r \, \widetilde{G}_{- r} \widetilde{G}_r 
    - \dfrac{1}{12} \, L_0 + \dfrac{17 c}{2880} 
    & (\mathrm{NS}) , \\[1.2ex]
    2 \sum \limits_{\substack{r > 0 \\ r \in \mathbb{Z}}} 
    r \, \widetilde{G}_{- r} \widetilde{G}_r 
    + \dfrac{1}{6} \, L_0 - \dfrac{7 c}{720} 
    & (\mathrm{R}) . 
  \end{cases}
\end{equation}

\subsubsection*{Spin-6: \texorpdfstring{$\bm{(\widetilde{T} (\partial \widetilde{G} \, \widetilde{G}))}$}{$(T (\partial G \, G))$} and \texorpdfstring{$\bm{(\partial^2 \widetilde{G} \, \partial \widetilde{G})}$}{$(\partial^2 G \, \partial G)$}}
For the spin-6 current, the contributions involving only $\widetilde{T}$ are already known from \cite{bazhanov1996,dymarsky2020,novaes2021}. 
We therefore record here only the zero modes of the fermionic contributions 
$(\partial^2 \widetilde{G} \, \partial \widetilde{G})$ and $(\widetilde{T} (\partial \widetilde{G} \, \widetilde{G}))$, which are needed to determine the zero mode of $\widetilde{J}_6$. 
The required OPE coefficients are obtained straightforwardly from the super-Virasoro OPEs, and we quote only the final results below. 
\begin{align}
  (\partial^2 \widetilde{G} \, \partial \widetilde{G})_0 &=
  \begin{cases}
    - 2 \sum \limits_{\substack{r > 0 \\ r \in \mathbb{Z} + \frac{1}{2}}}
    r^3 \widetilde{G}_{- r} \widetilde{G}_r
    - \dfrac{7}{480} \, L_0
    + \dfrac{457 c}{241920}
    & (\mathrm{NS}) , \\[1.2ex]
    - 2 \sum \limits_{\substack{r > 0 \\ r \in \mathbb{Z}}}
    r^3 \widetilde{G}_{- r} \widetilde{G}_r
    + \dfrac{1}{60} \, L_0
    - \dfrac{61 c}{30240}
    & (\mathrm{R}) , 
  \end{cases}
\end{align}
\begin{equation}
  (\widetilde{T} (\partial \widetilde{G} \, \widetilde{G}))_0 = 
  \begin{cases}
    \sum \limits_{n = 1}^{\infty} 
    \left[ \widetilde{L}_{- n} (\partial \widetilde{G} \, \widetilde{G})_n 
    + (\partial \widetilde{G} \, \widetilde{G})_{- n} \widetilde{L}_n \right] 
    - \dfrac{1}{12} \, L_0^2 + \dfrac{3 (c + 4)}{320} \, L_0 
    - \dfrac{c (119 c + 1196)}{483840} 
    & (\mathrm{NS}) , \\[1.2ex]
    \sum \limits_{n = 1}^{\infty} 
    \left[ \widetilde{L}_{- n} (\partial \widetilde{G} \, \widetilde{G})_n 
    + (\partial \widetilde{G} \, \widetilde{G})_{- n} \widetilde{L}_n \right] 
    + \dfrac{1}{6} \, L_0^2 - \dfrac{4 c + 49}{240} \, L_0 
    + \dfrac{c (49 c + 22381)}{120960} 
    & (\mathrm{R}) . 
  \end{cases}
\end{equation}